\documentclass[11pt]{article}
\pdfoutput=1
\usepackage{graphicx}
\usepackage{hyperref}
\usepackage{epsfig}
\usepackage{amsmath}
\usepackage{amsthm}
\usepackage{amssymb}
\usepackage{multirow}
\usepackage{color}
\usepackage[nosort]{cite}
\usepackage{hyperref}
\usepackage[fontsize=11.45pt]{scrextend}
\usepackage{braket,mathrsfs,slashed}
\usepackage{float}
\usepackage{setspace}
\usepackage[caption=false]{subfig}
\usepackage[export]{adjustbox}

\newlength{\abstractwidth}
\newcommand{\be}{\begin{equation}}
	\newcommand{\ee}{\end{equation}}

\renewcommand{\title}[1]{\vbox{\center\bf{\Large{#1}}}\vspace{5mm}}
\renewcommand{\author}[1]{\vbox{\center#1}\vspace{5mm}}
\newcommand{\address}[1]{\vbox{\center\em#1}}
\newcommand{\email}[1]{\vbox{\center\tt#1}\vspace{5mm}}

\usepackage{dsfont}

\usepackage[utf8]{inputenc}
\usepackage{rotating}
\usepackage{tikz-feynman}

\usepackage{dsfont}
\usepackage[paper=a4paper,margin=1.5cm]{geometry}
\usepackage[utf8]{inputenc}
\usepackage{rotating}
\usepackage{soul}

\usepackage{mathrsfs} 
\usepackage{bbm}
\usepackage{etoolbox} 
\usepackage{multirow}

\allowdisplaybreaks

\hypersetup{pdfstartview=FitV,colorlinks=true,linkcolor=midblue,citecolor=midblue,filecolor=midblue,urlcolor=midblue}

\definecolor{midblue}{rgb}{0,0,0.5}

\begin{document}
	
	\newgeometry{top=3.1cm,bottom=3.1cm,right=2.4cm,left=2.4cm}
	
	\begin{titlepage}
		\begin{center}
			\hfill \\
			\vskip 0.5cm

			\title{Remarks on ghost resonances}

			\author{\large Luca Buoninfante}
			
			\address{High Energy Physics Department, Institute for Mathematics, Astrophysics,
				\\and Particle Physics, Radboud University, Nijmegen, The Netherlands}
			\email{\rm \href{mailto:luca.buoninfante@ru.nl}{luca.buoninfante@ru.nl}}

		\end{center}

		\begin{abstract}
		In this paper we study various aspects of ghost       resonances: the resummation that leads to the dressed propagator, the poles locations, the analytic continuation into the second Riemann sheet and the spectral representations in both first and second sheets. In particular, we show that for real masses above the multiparticle threshold the ghost propagator has a pair of complex conjugate poles in the first sheet, unlike the case of an ordinary unstable resonance which has no pole in the first sheet but a complex conjugate pair in the second sheet. Mathematical and physical implications of this feature are discussed. We also clarify an important point regarding the two absorptive contributions of a ghost propagator in the narrow-width approximation. Furthermore, we argue that finite-time quantum field theories are needed to consistently derive the dressed ghost propagator and capture the true physical properties of ghost resonances. Throughout the work, different prescriptions to define the ghost propagator on the real axis are considered: Feynman, anti-Feynman and fakeon prescriptions.
		\end{abstract}
		
	\end{titlepage}
	
	{
		\hypersetup{linkcolor=black}
		\tableofcontents
	}

	\baselineskip=17.63pt

	

	\newpage
	
	\section{Introduction}
	
	Ghosts are fields whose kinetic operators and propagators have an opposite sign to that of ordinary fields. They appear naturally in higher-derivative field theories, where they are typically responsible for improving the ultraviolet behavior of loop diagrams through a faster fall-off of the propagator.	
	
	A famous example of a theory containing a ghost field traces back to more than fifty years ago when Lee and Wick~\cite{Lee:1969fy,Lee:1970iw} (see also~\cite{Cutkosky:1969fq,Coleman:1969xz,Nakanishi:1971jj,Nakanishi:1972wx,Nakanishi:1972pt,Boulware:1983vw}) proposed a four-derivative extension of quantum electrodynamics in order to obtain a finite quantum theory describing the interaction between photons and electrons. 
	Although the original proposal lost its relevance when it was understood that quantum electrodynamics could be embedded into the Standard Model, recently there have been new interests in Lee-Wick models~\cite{Grinstein:2008bg,Grinstein:2007mp,Carone:2008iw,Anselmi:2017yux,Anselmi:2017lia,Anselmi:2018kgz,Donoghue:2018lmc}.
	
	A more relevant and interesting example of four-derivative field theory is Quadratic Gravity~\cite{Stelle:1976gc,Tomboulis:1977jk,Julve:1978xn,Fradkin:1981iu,Avramidi:1985ki,Antoniadis:1986tu,Johnston:1987ue,Salvio:2014soa,Salvio:2018crh,Donoghue:2018izj,Donoghue:2019ecz,Donoghue:2021meq,Anselmi:2017ygm,Anselmi:2018ibi,Anselmi:2018tmf,Anselmi:2018bra,Donoghue:2021cza,Holdom:2021hlo,Holdom:2021oii,Piva:2023bcf,Buoninfante:2023ryt,Buccio:2024hys} whose Lagrangian contains all possible four-dimensional operators in addition to the Einstein-Hilbert term, in particular the Ricci scalar square and the Weyl square. In this case the ghost is a massive spin-two field coming from the Weyl square.  This theory is unique in the sense that it is strictly renormalizable in four spacetime dimensions~\cite{Stelle:1976gc} as any other quantum field theory (QFT) in the Standard Model.
	
	QFTs containing ghosts are sometimes considered unphysical due to possible pathologies such as Hamiltonian instability and unitarity violation~\cite{Ostrogradsky:1850fid,Woodard:2015zca,Buoninfante:2022ykf}. Indeed, by the late 1980s, Quadratic Gravity had been largely forgotten, mainly because people were afraid of ghosts~\cite{Nakanishi:1971jj,Nakanishi:1972wx,Nakanishi:1972pt,Boulware:1983vw,Johnston:1987ue}. It should also be mentioned that other (less conventional) approaches to quantum gravity began to gain popularity during exactly those years and the idea of completing Einstein's general relativity with renormalizable quadratic-curvature terms was abandoned. 
	
	However, especially in the last decade,  higher-derivative field theories, together with the old lore of quantizing gravity in the framework of perturbative QFT, have been going through a renaissance. Indeed, new studies on Lee-Wick theories~\cite{Grinstein:2008bg,Grinstein:2007mp,Carone:2008iw,Anselmi:2017yux,Anselmi:2017lia,Anselmi:2018kgz,Donoghue:2018lmc}, Quadratic Gravity~\cite{Salvio:2014soa,Salvio:2018crh,Donoghue:2018izj,Donoghue:2019ecz,Donoghue:2021meq,Anselmi:2017ygm,Anselmi:2018ibi,Anselmi:2018tmf,Anselmi:2018bra,Donoghue:2021cza,Holdom:2021hlo,Holdom:2021oii,Piva:2023bcf,Buoninfante:2023ryt,Buccio:2024hys,Kuntz:2024rzu}, and classical dynamical systems containing ghosts~\cite{Deffayet:2021nnt,Deffayet:2023wdg,ErrastiDiez:2024hfq} have been carried out, and new ideas to make sense of ghosts have been proposed.

	Ghost particles behave in many ways very differently from ordinary particles. For instance, in standard QFTs without ghosts, quantum effects can turn a stable particle into an unstable resonance whose mass and width are mathematically related to the real and imaginary parts of a complex pole appearing in the second Riemann sheet of the dressed propagator~\cite{Landshoff:1963nzy,Eden:1966dnq,Brown:1992db,Coleman:2018mew}. The absence of poles in the first sheet guarantees that the unstable particle is no longer an asymptotic state and can also be projected out of the set of intermediate states consistently with unitarity via the so-called Veltman projection~\cite{Veltman:1963th}. However, if the particle is a ghost, then the pole does not move to the second sheet, but a pair of complex conjugate poles appears in the first sheet~\cite{Lee:1969fy,Lee:1970iw,Coleman:1969xz,Antoniadis:1986tu}. Despite this unusual feature,  the original idea proposed by Lee and Wick~\cite{Lee:1969fy,Lee:1970iw} was that a ghost with complex energy cannot appear on-shell because of energy conservation (see also~\cite{Coleman:1969xz}). More recently, the possibility to have a ghost particle that decays and disappears from the set of asymptotic and intermediate states was considered in~\cite{Grinstein:2008bg,Donoghue:2019fcb,Donoghue:2021cza}. On the contrary, Refs.~\cite{Kubo:2023lpz,Kubo:2024ysu} pointed out that the presence of complex poles in the first sheet implies that a ghost particle actually contributes to the set of asymptotic states with complex masses and that it can be produced by collisions between ordinary particles and does not decay.

	\subsubsection*{Aim of this work}

	In this paper we intend to discuss several aspects of ghosts and shed light on various features of ghost resonances that can sometimes give rise to confusion. From a mathematical point of view, we will study in detail the resummation of self-energies leading to the dressed propagator, the locations of the poles, the analytic continuation into the second Riemann sheet, and the spectral representations of the propagator in both sheets. We will then analyse the physical implications and try to grasp some of the key properties that distinguish a ghost resonance from an ordinary unstable resonance, including the question of whether the decay occurs.

	The work is organized as follows.

	\begin{description}
		
		\item[Sec.~\ref{sec:model}:] We consider a two-derivative scalar QFT model whose interaction term could in principle describe the decay channel of a ghost into two ordinary particles. We introduce the Feynman, anti-Feynman and fakeon prescriptions for the ghost propagator. 
		Furthermore, we compute the dressed propagator by resumming the one-loop self-energies and perform a very detailed analysis of the poles locations, first in the ordinary case without ghost and then in the case with ghost. We explicitly determine the two physically distinct absorptive parts of the propagator and clarify an important point about (a)causal propagation and arrows of time in the narrow-width approximation. 
		
		\item[Sec.~\ref{sec:general-analysis}:] Using model-independent properties of the propagator and without relying on perturbation theory, we perform the analytic continuation into the second Riemann sheet and define the spectral representations in both sheets. We clarify the meanings of the pole-like structure of the propagator in both sheets when working in the narrow-width approximation.

		\item[Sec.~\ref{sec:discuss}:] We make further remarks on the difference between an unstable ordinary resonance and a ghost resonance for the Feynman, anti-Feynman and fakeon prescriptions, and comment on unitarity, asymptotic states and decay. We also note that the presence of complex poles in the first sheet of a ghost propagator does not allow the use of analyticity to resum the geometric series in the region around the peak. We discuss mathematical and physical implications of this feature. In particular, we point out towards the need to work with a QFT formulated in a finite  interval of time to consistently compute the dressed propagator and capture the true physical properties of a ghost resonance.
		
		\item[Sec.~\ref{sec:concl}:] We summarize our results and draw the conclusions.

	\end{description}

	\paragraph{Conventions.} Throughout this work we work in Natural units, $\hbar=1=c,$ and adopt the mostly plus convention for the metric signature, $(\eta_{\mu\nu})=\text{diag}(-1,+1,+1,+1).$ The latter implies that $p^2=p_\mu p^\mu$ is negative (positive) for a time-like (space-like) vector $p^\mu.$ Moreover, an ordinary scalar kinetic term has a minus sign in front, i.e. $-\frac{1}{2}\partial_\mu \chi \partial^\mu\chi=\frac{1}{2}(\partial_0\chi)^2-\frac{1}{2}\partial_i\chi\partial^i\chi$, while a ghost kinetic term has a plus sign, i.e. $+\frac{1}{2}\partial_\mu \phi \partial^\mu\phi=-\frac{1}{2}(\partial_0\phi)^2+\frac{1}{2}\partial_i\phi\partial^i\phi $.

	\section{A field theory model}\label{sec:model}
	
	Let us consider the following Lagrangian
	\begin{equation}
		\mathcal{L}= -\frac{1}{2}\left(\partial_\mu\chi\partial^\mu\chi +\mu^2\chi^2\right) - a\frac{1}{2} \left(\partial_\mu\phi\partial^\mu \phi+m^2\phi^2\right)-\frac{g}{2} \phi\chi^2\,,
		\label{lagrangian}
	\end{equation}
	where $\chi$ is an ordinary scalar field with mass $\mu$, while $\phi$ is an ordinary scalar field if $a=1$ or a ghost if $a=-1,$ with mass $m.$ The interaction term with coupling constant $g$ can in principle allow for the decay of a $\phi$-particle into two $\chi$-particles if the inequality $m> 2\mu$ holds true. We assume that $m$ is already the renormalized physical mass in agreement with the renormalization condition that will be chosen in sec.~\ref{sec:resummed}.

	\subsection{Tree-level propagator and unitarity}\label{sec:tree-unitarity}

	The unprescribed tree-level propagators of the fields $\chi$ and $\phi$ are respectively given by
	\begin{eqnarray}
		G_\chi(-p^2) \!\!\!&=&\!\!\! \frac{-i}{p^2+\mu^2}\,\, \equiv \,\,  \vcenter{\hbox{\includegraphics[scale = 0.35]{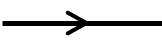}}}\,, 	\label{propagator-phi}\\[1.5mm]
		G_{\phi,a}(-p^2)\!\!\!&=&\!\!\! \frac{-ia}{p^2+m^2}\,\, \equiv \,\,  \vcenter{\hbox{\includegraphics[scale = 0.35]{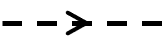}}}\,,
		\label{propagator-chi}
	\end{eqnarray}
	where the arrows represent the energy flows. To fully define the propagator we should also specify what prescription is used to handle the pole singularities, i.e. how to move the poles away from the real axis. The $\chi$-propagator is prescribed with the ordinary Feynman shift, i.e.
	\begin{eqnarray}
		G_\chi(-p^2) \rightarrow  \frac{-i}{p^2+\mu^2-i\varepsilon}\,,
	\end{eqnarray}
	where $\varepsilon \rightarrow 0^+.$ On the other hand, if the $\phi$-propagator is ghost-like, then different types of prescriptions have been considered in the literature. A general expression which compactly collects various prescriptions is given by
	\begin{eqnarray}
	G_{\phi,a}(-p^2)\rightarrow \frac{-ia}{2}\left[\frac{1}{p^2+m^2-ib\epsilon}+\frac{1}{p^2+m^2-ic\epsilon} \right]\,,
	\end{eqnarray}
	where $\epsilon\rightarrow 0^+$ and we have introduced the parameters $b=\pm 1$ and $c=\pm 1.$ 
	\begin{itemize}
		
		\item If $b=c=1$, we have the \textit{Feynman} prescription~\cite{Lee:1969fy,Coleman:1969xz,Holdom:2021hlo}:  the tree-level propagator is continued onto the real axis from above and positive (negative) energies propagate forward (backward) in time, i.e. we have a time-ordered product in position space.
		
		\item If $b=c=-1,$ we have the \textit{anti-Feynman} prescription~\cite{Donoghue:2019ecz,Donoghue:2019fcb,Donoghue:2021cza}:  the tree-level propagator is continued onto the real axis from below and positive (negative) energies propagate backward (forward) in time, i.e. we have an anti-time-ordered product in position space. 	
		
		\item If $b=-c=\pm 1,$ we have the \textit{fakeon} prescription~\cite{Anselmi:2017ygm,Anselmi:2018bra,Piva:2023bcf}: the tree-level propagator is an average of Feynman and anti-Feynman propagators in momentum space, i.e. we have an average of time-ordered and anti-time-ordered products in position space.
		
	\end{itemize}

	Unitarity, i.e. the conservation of quantum probabilities, imposes constraints on the signs of  $a$, $b$ and $c.$ In terms of the $S$-matrix the unitarity relation is $S^\dagger S=\mathds{1},$ and if we introduce the transfer matrix $T$ defined by $S=\mathds{1}+i T,$ we obtain the operator form of the optical theorem $i(T^\dagger - T)=T^\dagger T,$ whose diagonal elements describe elastic scattering processes. For example, given an asymptotic state $\left|\psi \right\rangle $ we can write
	\begin{eqnarray}
		2\text{Im}\left[\left\langle \psi \right| T \left| \psi \right\rangle  \right] =\sum_{n}\sigma_n \left| \left\langle n \right| T \left| \psi \right\rangle \right|^2\,,  
		\label{optical-theorem}
	\end{eqnarray}
	where on the right-hand side we have inserted the completeness relation $\mathds{1}=\sum_n \sigma_n \left| n \right\rangle \left\langle n \right|$ with the constants $\sigma_n$ defined by $\langle n |m \rangle=\sigma_n\delta_{nm}.$ The form of the optical theorem in eq.~\eqref{optical-theorem} is still schematic because the integral over the phase space is hidden in the sum $\sum_n$ and the Dirac deltas that take into account the momentum conservation have not been factored out. Anyway, these additional details are not needed for our discussion and so we omit them.
	
	It is important to emphasize that in general the requirement of probability conservation does not say anything about the sign of the imaginary part of an amplitude or about the signs of the norms, in fact unitarity can be satisfied even if some of the coefficients $\sigma_n$ are negative. However, unitarity imposes constraints on the relative signs of  $a$, $b,$ $c$ and $\sigma_n$. For example, if we consider the simplest case of a tree-level scattering process $\chi\chi\rightarrow \chi\chi$ mediated by the $\phi$-field, the left-hand side of eq.~\eqref{optical-theorem} reads
	\begin{equation}
		\begin{aligned}
		2\text{Im}\left[\left\langle \psi \right| T \left| \psi \right\rangle  \right] =&\, 2\text{Im} \left[(-i)(-ig)^2 \frac{-i a}{2}\left(\frac{1}{p^2+m^2-ib\epsilon}+\frac{1}{p^2+m^2-ic\epsilon}\right) \right]\\[1mm]
		=&\,\pi g^2\, a(b+c) \,\delta(p^2+m^2)\,,
		\end{aligned}
	\end{equation}
	where in this case the two-particle state $\left| \psi \right\rangle = \left| p_1,p_2 \right\rangle=\left| p_3,p_4 \right\rangle$ is the same for both ingoing and outgoing states since the process is elastic, and $p=p_1+p_2=p_3+p_4$ is the total ingoing or outgoing momentum. Let us discuss various cases in which unitarity can be satisfied.
	\begin{itemize}
		
		\item In the standard scenario of a $\phi$-field with $a=1$ the norms of the physical states are positive, i.e. $\sigma_n>0.$ Then, the optical theorem requires the Feynman prescription with $b=c=1$. 
		
		
	\end{itemize}

	\noindent If the $\phi$-field is a ghost, i.e. $a=-1,$ we have at least three options that can satisfy unitarity:

	\begin{itemize}

		\item If $b=c=1,$ the imaginary part of the amplitude cannot be positive, i.e. $\text{Im}\left[\left\langle \psi \right| T \left| \psi \right\rangle  \right]\leq 0.$ This means that~\eqref{optical-theorem} can be valid if and only if the norms of states containing an odd number of ghost particles are negative, i.e. $\sigma_{2n+1}<0.$ In this case we call the $\phi$-field \textit{Feynman ghost}.
		
		\item If $b=c=-1,$ the imaginary part of the amplitude cannot be negative, i.e.~$\text{Im}\left[\left\langle \psi \right| T \left| \psi \right\rangle  \right]\geq 0.$ This means that~\eqref{optical-theorem} can be valid if and only if the norms of the ghost states are positive, i.e. $\sigma_n>0.$ In this case we call the $\phi$-field  \textit{anti-Feynman ghost}.
		
		\item If $b=-c=\pm 1,$ the imaginary part of the amplitude vanishes, i.e. $\text{Im}\left[\left\langle \psi \right| T \left| \psi \right\rangle  \right]= 0.$ This means that~\eqref{optical-theorem} can be valid if and only if the right-hand side of the optical theorem is also equal to zero. This can be achieved by projecting out all the ghost states from the unitarity sum, thus the ghost is converted into a purely virtual off-shell degree of freedom.  In this case we call the $\phi$-field  \textit{fakeon ghost}.
		
	\end{itemize}

	We have only discussed tree-level unitarity, but higher loops have also been studied in the literature. For a Feynman ghost, loop integrals in perturbation theory can be handled in the standard way since analyticity properties still hold (at least perturbatively), thus Lorentzian and Euclidean results are linked by the usual Wick rotation. As for theories with anti-Feynman or fakeon ghosts, the standard analyticity is lost already at the perturbative level and the usual Wick rotation is no longer available. Alternative rules for (non-analytic) contour deformations have been proposed to compute loop integrals and prove unitarity at all orders in perturbation theory in the presence of fakeons~\cite{Anselmi:2018kgz,Anselmi:2021hab}, and partial results up to one loop also exist for theories containing anti-Feynman ghosts~\cite{Donoghue:2019fcb}. However, it is still unclear whether anti-Feynman and fakeon ghosts can be consistently described within the operator formalism of QFT~\cite{Kubo:2023lpz,Holdom:2024cfq}, therefore future work is certainly needed to address this question. 
	
	Here we will be interested in resumming all the self-energy contributions and discussing  mathematical and physical features of the non-perturbative dressed propagator. We will mainly focus on the Feynman and anti-Feynman prescriptions for the ghost propagator, but also comment on the fakeon ghost towards the end. 
	
	\paragraph{Notations.} We will often consider the $\phi$-propagator as a function of complex momentum square $-p^2=z\in \mathbb{C},$ in particular at tree level we will write 
	\begin{equation}
	G_{\phi,a}(z)=\frac{ia}{z-m^2}\,.
	\end{equation}
	We will denote the real part of $z$ as $s= {\rm Re}[z].$ Then, the Feynman ($b=1$) and anti-Feynman ($b=-1$) tree-level propagators as a function of the real momentum square $s$ will be defined as
	\begin{equation}
		G_{\phi,a}(s,b\epsilon)=\frac{ia}{s-m^2+ib\epsilon}\,.
	\end{equation}

	\subsection{Dressed propagator}\label{sec:resummed}
	
	We want to compute the quantum corrections to the $\phi$-propagator by resumming the one-loop self-energies to all orders in perturbation theory. 
	
	\paragraph{Self-energy.} The self-energy $\Sigma(z)$ at one-loop as a function of $-p^2=z\in \mathbb{C}$ is given by
	\begin{equation}
		\begin{aligned}
		i\Sigma(z)\equiv  \,\vcenter{\hbox{\includegraphics[scale = 0.23]{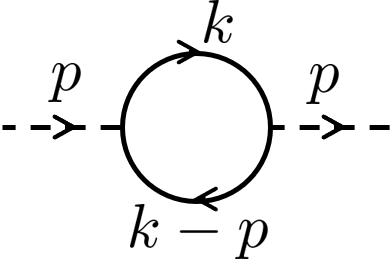}}} =&\,  (-ig)^2\int \frac{{\rm d}^4k}{(2\pi)^4} \frac{-i}{k^2+\mu^2-i\varepsilon} \frac{-i}{(k-p)^2+\mu^2-i\varepsilon}\\[1mm]
		=&\,i\frac{g^2}{16\pi^2}\left[2-\log\left(\frac{\mu^2}{\Lambda^2}\right)-2\sqrt{\frac{4\mu^2-z}{z}}\arctan\sqrt{\frac{z}{4\mu^2-z}}\right]\,,
		\end{aligned}
		\label{self-energy}
	\end{equation}
	where $\Lambda$ is the renormalization scale in the cut-off scheme, and only the Feynman shifts ($\varepsilon\rightarrow 0^+$) appear because only the $\chi$-propagators are attached to the internal lines. This means that the integral~\eqref{self-energy} gives the standard result for the self-energy, in particular if we take the momentum square to be real and time-like, i.e. $z\rightarrow s=-p^2>0,$ we can evaluate the real and imaginary parts of $\Sigma(s\pm i\epsilon)$: 
	\begin{eqnarray}
		{\rm Re}\left[\Sigma(s)\right]=\frac{g^2}{16\pi^2}\left[ 2-\log\left(\frac{\mu^2}{\Lambda^2}\right)+\sqrt{1-\frac{4\mu^2}{s}}\log\left(\frac{\sqrt{1-4\mu^2/s}-1}{\sqrt{1-4\mu^2/s}+1}\right) \right]
		\label{real-self-energy}
	\end{eqnarray}
	and
	\begin{equation}
		{\rm Im}\left[\Sigma(s\pm i\epsilon)\right]=\pm\frac{g^2}{16\pi}\sqrt{1-\frac{4\mu^2}{s}}\,\theta\left(s-4\mu^2\right)\equiv \pm \gamma(s)\,.
		\label{imag-self-en}
	\end{equation}
	The sign of the imaginary part depends on whether we continue $\Sigma(z)$ onto the real axis from above $(z=s+i\epsilon)$ or from below $(z=s-i\epsilon)$. We will consider $\Sigma(s+i\epsilon)$ in the expression of the dressed propagator\footnote{This choice follows automatically if the dressed propagator as a function of real momentum can be defined by analytically continuing onto the real axis from above; this happens for a Feynman ghost. However, if this cannot be done due to some lack of analyticity, then the choice $\Sigma(s+i\epsilon)$ must be taken as an additional assumption; this happens for the anti-Feynman and fakeon ghosts, as will become clearer later.\label{choice-imag-self}} since this is the one that gives a positive absorptive contribution above the multiparticle threshold, as we will see later.
	
	It is important to emphasize that eq.~\eqref{imag-self-en} refers to the imaginary part of the self-energy in the first Riemann sheet. However, the signs are opposite in the second sheet. We can find the expression of the self-energy in the second sheet, which we call $\Sigma^{II}(z)$, by analytically continuing the function $\Sigma(z)$ from the first sheet downward through the branch cut, i.e.
	\begin{equation}
		\Sigma^{II}(s-i\epsilon)=\Sigma(s+i\epsilon)=\Sigma(s-i\epsilon)+2i\gamma(s)\,,
	\end{equation}
	where we have used $\gamma(s)=\frac{1}{2i}[\Sigma(s+i\epsilon)-\Sigma(s-i\epsilon)].$ By taking the complex conjugate and using $\Sigma^{*}(z)=\Sigma(z^*),$ we can also write
	\begin{equation}
		\Sigma^{II}(s+i\epsilon)=\Sigma(s+i\epsilon)-2i\gamma(s)\,.
	\end{equation}
	Then, by analytically continuing $\gamma(s)$ to the complex plane, we obtain the self-energy in the second Riemann sheet:
	\begin{equation}
		\Sigma^{II}(z)=\Sigma(z)-2i\gamma(z)\,,
		\label{self-second}
	\end{equation}
	where $\gamma(s\pm i\epsilon)=\pm \gamma(s).$ The imaginary part for real and time-like momentum is given by
	\begin{equation}
		{\rm Im}\left[\Sigma^{II}(s\pm i\epsilon)\right]=\mp \gamma(s)\,,
		\label{imag-self-second}
	\end{equation}
	which has the opposite sign to the counterpart in the first sheet~\eqref{imag-self-en}. This sign difference will be crucial to determine the location of the poles of the dressed propagator.

	\paragraph{Resummation.} The dressed (also called resummed) $\phi$-propagator $\bar{G}_{\phi,a}(z)$ as a function of complex momentum $-p^2=z\in \mathbb{C}$ can be computed as follows
	\begin{eqnarray}
		\bar{G}_{\phi,a}(z)\!\!\!&\equiv&\!\!\! \vcenter{\hbox{\includegraphics[scale = 0.23]{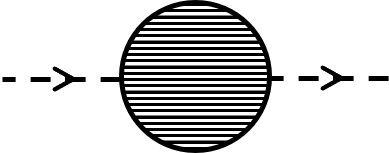}}} \nonumber \\[1mm]
		&=&\!\!\!  \vcenter{\hbox{\includegraphics[scale = 0.23]{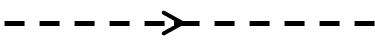}}}\,+\, \vcenter{\hbox{\includegraphics[scale = 0.23]{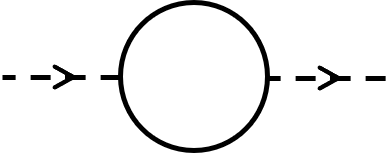}}}\,+\,\vcenter{\hbox{\includegraphics[scale = 0.23]{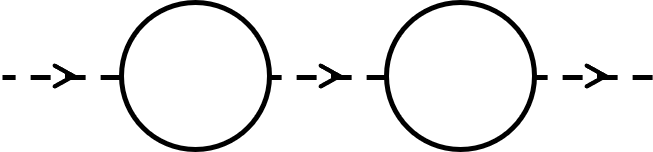}}}\,+\cdots\nonumber\\[1mm]
		&=&\!\!\!  G_{\phi,a}(z)+ G_{\phi,a}(z)i\Sigma(z) G_{\phi,a}(z)+ G_{\phi,a}i\Sigma(z) G_{\phi,a}i\Sigma(z)G_{\phi,a}(z)+\cdots \nonumber \\[1mm]
		&=&\!\!\!  G_{\phi,a}(z)\sum_{n=0}^\infty \left[i\Sigma(z)G_{\phi,a}(z) \right]^n\,.
		\label{geometric-series}
	\end{eqnarray}
	The requirement of convergence of the geometric series demands
	\begin{eqnarray}
		\left|i\Sigma(z)G_{\phi,a}(z) \right|<1 \qquad \Leftrightarrow \qquad \frac{\Sigma^*(z)\Sigma(z)}{(z-m^2)(z^*-m^2)}<1
	\end{eqnarray}
	In the region around $z\simeq m^2$ the last inequality can be violated if $\text{Im}[\Sigma(z= m^2)]\neq 0,$ and the geometric series may fail to converge. In such a case, the usual procedure is to first sum the geometric series far from $z\simeq m^2$, and then use analyticity to continue the result in the region around $z\simeq m^2$~\cite{Veltman:1963th}. Assuming that this can be done, we can write
	\begin{eqnarray}
		\bar{G}_{\phi,a}(z)=\frac{G_{\phi,a}(z)}{1-i\Sigma(z)G_{\phi,a}(z)}=\frac{ia}{z-m^2+a\Sigma(z)}\,.
		\label{resummed-propag}
	\end{eqnarray}
	Note that, rigorously speaking, the resummation can be trusted for any complex momentum if the dressed propagator~\eqref{resummed-propag} is analytic everywhere or, to be more precise, if $\bar{G}_{\phi,a}(z)$ does not have any singularities in the first Riemann sheet of the complex $z$ plane (it can have branch cuts). As we will explain in sec.~\ref{sec:inval-resum}, this is not guaranteed when $a=-1$.
	
	For real momentum, i.e. $z\rightarrow s,$ we have to distinguish between Feynman and anti-Feynman prescriptions, thus eq.~\eqref{resummed-propag} becomes
	\begin{eqnarray}
		\bar{G}_{\phi,a}(s,b\epsilon)=\frac{G_{\phi,a}(s,b\epsilon)}{1-i\Sigma(s+i\epsilon)G_{\phi,a}(s,b\epsilon)}=\frac{ia}{s-m^2+ib\epsilon+a\Sigma(s+i\epsilon)}\,.
		\label{resummed-propag-feyn-anti-feyn}
	\end{eqnarray}

	\paragraph{Remark.} Note that the presence of $b\epsilon$ in the argument of the dressed propagator means that the tree-level propagator in the geometric series is prescribed with $s+ib\epsilon.$ On the other hand, the self-energy $\Sigma(z)$  is always continued onto the real axis from above, i.e. with $z=s+i\epsilon;$ see the discussion after eq.~\eqref{imag-self-en}. This is one way to understand that the anti-Feynman prescription is not analytic already at the perturbative level. Indeed, for Feynman $(b=1)$ and anti-Feynman $(b=-1)$ prescriptions we have
	\begin{eqnarray}
	\bar{G}_{\phi,a}(s,\epsilon)=\bar{G}_{\phi,a}(s+i\epsilon)\qquad \text{and} \qquad \bar{G}_{\phi,a}(s,-\epsilon)\neq \bar{G}_{\phi,a}(s-i\epsilon)\,,
	\end{eqnarray}
	respectively. In particular, the latter relation means that the function $\bar{G}_{\phi,a}(s,-\epsilon)$ cannot be defined as the analytic continuation of $\bar{G}_{\phi,a}(z)$ with $z=s-i\epsilon$ and $\epsilon\rightarrow 0^+.$ This lack of analyticity is present even before resumming the geometric series, that is, already perturbatively, and occurs for the anti-Feynman ghost; this also clarifies the observation in footnote~\ref{choice-imag-self}. In sec.~\ref{sec:discuss}, we will compare Feynman and anti-Feynman prescriptions and further comment on the lack of analyticity, in particular we will distinguish between perturbative and non-perturbative breaking of analyticity.
	
	\subsubsection{Two absorptive contributions}
	
	The absorptive part of the propagator is
	\begin{equation}
		{\rm Im}\left[i\bar{G}_{\phi,a}(s,b\epsilon)\right]=\frac{ab\epsilon+\gamma(s)}{\left(s-m^2+a{\rm Re}\left[\Sigma(s)\right]\right)^2+\left(b\epsilon+a\gamma(s)\right)^2}\,,
		\label{imag-iG}
	\end{equation}
	which is usually related to the spectral density. The two contributions in the numerator have two different and equally important physical meanings, as we will now explain.
	
	Let us write 
	\begin{eqnarray}
		2{\rm Im}\left[i\bar{G}_{\phi,a}(s,b\epsilon)\right]=2A_a(s,b\epsilon)+2B_a(s,b\epsilon)\,,
		\label{2imag-iG}
	\end{eqnarray}
	where we have defined
	\begin{eqnarray}
		A_a(s,b\epsilon)\!\!\!&\equiv&\!\!\! \bar{G}_{\phi,a}(s,b\epsilon)\,ab\epsilon \,\bar{G}_{\phi,a}^*(s,b\epsilon)\,,\label{A-function} \\[1mm]
		B_a(s,b\epsilon)\!\!\!&\equiv &\!\!\!\bar{G}_{\phi,a}(s,b\epsilon)\,\gamma(s)\, \bar{G}_{\phi,a}^*(s,b\epsilon)\,.\label{B-function}
	\end{eqnarray}
	The function $A_a(s,b\epsilon)$ takes into account processes where the $\phi$-particle appears on-shell and its propagation is characterized by a pole on the real axis. On the other hand, $B_a(s,b\epsilon)$ describes processes above the multiparticle threshold, for example in the standard case ($a=1=b$) it describes physical scenarios in which the $\phi$-particle is not alive and only the products of its decay can be detected.  Diagrammatically, these two contributions correspond to two different cuts~\cite{tHooft:1973wag}: the former corresponds to cutting the $\phi$-lines; the latter corresponds to cutting the self-energy. Let us show this explicitly.
	
	Defining the cut
	\begin{eqnarray}
	 \vcenter{\hbox{\includegraphics[scale = 0.23]{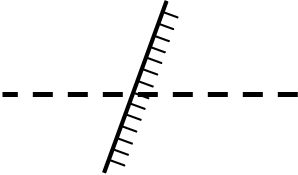}}}\,\equiv\, G_{\phi,a}(s,b\epsilon)+G^*_{\phi,a}(s,b\epsilon)\,,
	\end{eqnarray}
	we have
	\begin{eqnarray}
	\!\!\!\!\!\!\!\!\!\!	2A_a(s,b\epsilon)\!\!\!&=&\!\!\! \vcenter{\hbox{\includegraphics[scale = 0.23]{cut-line-right}}}\,+\,\vcenter{\hbox{\includegraphics[scale = 0.23]{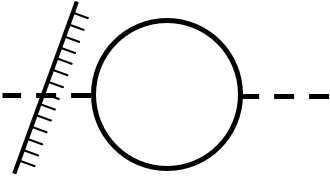}}}\,+\,\vcenter{\hbox{\includegraphics[scale = 0.23]{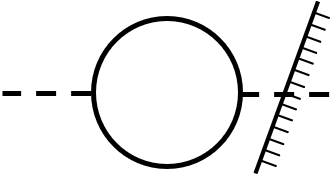}}}\,+\,\vcenter{\hbox{\includegraphics[scale = 0.23]{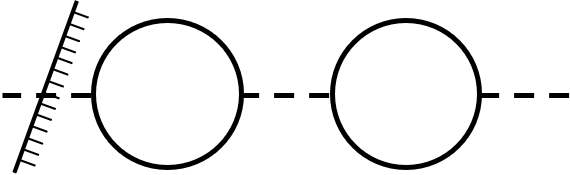}}}\,+\,\cdots \nonumber \\[1mm]
		&=&\!\!\! \vcenter{\hbox{\includegraphics[scale = 0.23]{cut-line-right}}}\,+\, \left(\vcenter{\hbox{\includegraphics[scale = 0.23]{cut-line-right}}}\right) \Big(\vcenter{\hbox{\includegraphics[scale = 0.23]{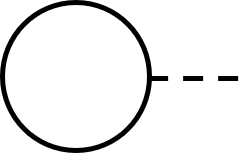}}}\,+\,\vcenter{\hbox{\includegraphics[scale = 0.23]{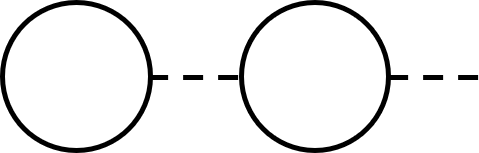}}}\,+\,\cdots\Big)^*+\cdots\nonumber\\[1mm]
		&=&\!\!\!\sum_{n=0}^{\infty}\big[i\Sigma(s) G_{\phi,a}(s,b\epsilon)\big]^n\big(G_{\phi,a}(s,b\epsilon)+ G^*_{\phi,a}(s,b\epsilon)\big) \sum_{k=0}^{\infty}\big[\big(i\Sigma(s) G_{\phi,a}(s,b\epsilon)\big)^*\big]^k\,,
		\label{A-function-diagram}
	\end{eqnarray}
	where the energy flow is taken from left to right, thus the part of the diagram to the right of the cut is complex conjugated. 	Then, using the relation
	\begin{equation}
		\begin{aligned}
	G_{\phi,a}(s,b\epsilon)+G^*_{\phi,a}(s,b\epsilon)&= G_{\phi,a}(s,b\epsilon)\frac{G_{\phi,a}(s,b\epsilon)+G^*_{\phi,a}(s,b\epsilon)}{G_{\phi,a}(s,b\epsilon)G^*_{\phi,a}(s,b\epsilon)}G^*_{\phi,a}(s,b\epsilon) \\[1mm]
	&= G_{\phi,a}(s,b\epsilon)\, 2 a b \,\epsilon \, G^*_{\phi,a}(s,b\epsilon)\,,
	\end{aligned}
	\end{equation}
	we get twice the expression in eq.~\eqref{A-function} for the absorptive term $2A_a(s,b\epsilon).$
	
	Similarly, if we define the cut
	\begin{eqnarray}
		\vcenter{\hbox{\includegraphics[scale = 0.23]{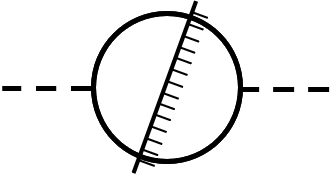}}}\,\equiv\, G_{\phi,a}(s,b\epsilon) 2\gamma(s) G^*_{\phi,a}(s,b\epsilon)\,,
	\end{eqnarray}
	we can show
	\begin{eqnarray}
		2B_a(s,b\epsilon)\!\!\!&=&\!\!\! \vcenter{\hbox{\includegraphics[scale = 0.23]{cut-line-bubble}}}\,+\,\vcenter{\hbox{\includegraphics[scale = 0.23]{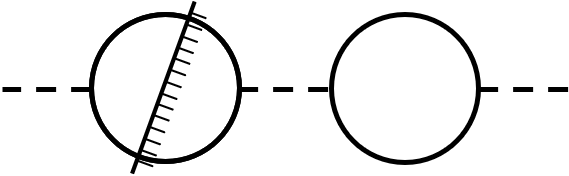}}}\,+\,\vcenter{\hbox{\includegraphics[scale = 0.23]{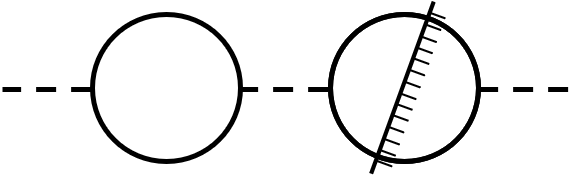}}}\,+\,\vcenter{\hbox{\includegraphics[scale = 0.23]{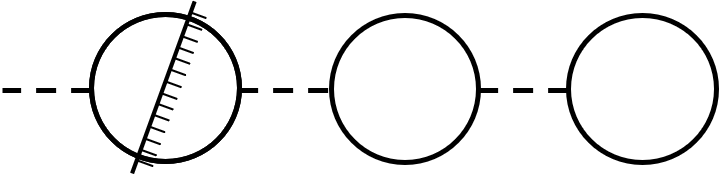}}}\,+\,\cdots \nonumber \\[1mm]
		&=&\!\!\! \vcenter{\hbox{\includegraphics[scale = 0.23]{cut-line-bubble}}}\,+\, \Big(\vcenter{\hbox{\includegraphics[scale = 0.23]{cut-line-bubble}}}\Big)\Big(\vcenter{\hbox{\includegraphics[scale = 0.23]{1-bubble}}}\,+\,\vcenter{\hbox{\includegraphics[scale = 0.23]{2-bubbles}}}\,+\,\cdots\Big)^*+\cdots\nonumber\\[1mm]
		&=&\!\!\!\sum_{n=0}^{\infty}\big[i\Sigma(s) G_{\phi,a}(s,b\epsilon)\big]^nG_{\phi,a}(s,b\epsilon)2\gamma(s)G^*_{\phi,a}(s,b\epsilon) \sum_{k=0}^{\infty}\big[\big(i\Sigma(s) G_{\phi,a}(s,b\epsilon)\big)^*\big]^k\,,\nonumber\\
		&&
		\label{B-function-diagram}
	\end{eqnarray}
	which is equal to twice the expression in eq.~\eqref{B-function}.
	
	Note that, while $B_a(s,b\epsilon)$ is always non-negative due to ${\rm Im}[\Sigma(s+i\epsilon)]=\gamma(s)\geq 0$,  the contribution $A_a(s,b\epsilon)$ is positive  only when $a=b=\pm 1$. The signs of the two absorptive parts of the dressed propagator will be relevant for the discussions in secs.~\ref{sec:remarks-limit} and~\ref{sec:stable-resonance}.

	\subsubsection{Narrow-width approximation} 
	
	Let us Taylor expand $\Sigma(z)$ around $z\simeq m^2$ and just above the real axis: 
	\begin{eqnarray}
		\Sigma(z)=\Sigma(m^2+i\epsilon)+\left.\frac{{\rm d}\Sigma(z)}{{\rm d}z}\right|_{z=m^2+i\epsilon}(z-m^2)+\mathcal{O}\left((z-m^2)^2\right)\,.
		\label{taylor-self}
	\end{eqnarray}
	If $m^2>4\mu^2$ the self-energy has both real and imaginary parts:
	\begin{eqnarray}
		\Sigma(m^2)\!\!\!&=&\!\!\!\text{Re}\left[\Sigma(m^2)\right]+i \,\text{Im}\left[\Sigma(m^2+i\epsilon)\right]\,,\\[1.5mm] 
		\left.\frac{{\rm d}\Sigma(z)}{{\rm d}z}\right|_{z=m^2}\!\!\!&=&\!\!\! \left.\text{Re}\left[\frac{{\rm d}\Sigma(z)}{{\rm d}z}\right]\right|_{z=m^2} +  i \left.\text{Im}\left[ \frac{{\rm d}\Sigma(z)}{{\rm d}z}\right]\right|_{z=m^2+i\epsilon}\,.
	\end{eqnarray}

	We can fix the renormalization conditions by requiring that the physical real mass of the $\phi$-field is equal to $m$ and that the residue of the resummed propagator at the pole $z=m^2$ below the multiparticle threshold is equal to $ia Z_R,$ where $Z_R>0$ is the wave-function renormalization constant. In formula, we have\footnote{Note that, in principle, we can choose any other equivalent renormalization condition without affecting the physics. For instance, if we started with a bare mass $m_0,$ then to define the renormalized mass we would need a non-zero value of $\text{Re}[\Sigma(m^2)]$ such that $m_0^2-a\text{Re}[\Sigma(m^2)]=m^2.$ Moreover, we could have also used a renormalization point different from $m^2$ as done in Ref.~\cite{Kubo:2024ysu} where a complex one was chosen.}
	\begin{equation}
		\text{Re}\left[\Sigma(m^2)\right]=0\,,\qquad \left.\text{Re}\left[\frac{{\rm d}\Sigma(z)}{{\rm d}z}\right]\right|_{z=m^2} =a^{-1}(Z_R^{-1}-1)\,.
		\label{renorm-cond}
	\end{equation}

	Having fixed the renormalization conditions up to order $\mathcal{O}(g^2),$ it follows that $Z_R^{-1}(z-m^2)$ is of order $\mathcal{O}(g^2)$. Therefore, we can write the dressed propagator in the region close to $z\simeq m^2$ as
	\begin{equation}
		\bar{G}_{\phi,a}(z)=\frac{iaZ_R}{z-m^2+iaZ_R\gamma(m^2)}+\mathcal{O}(g^4)\,,
		\label{dress-propag-g^2}
	\end{equation}
	where we have used ${\rm Im}[\Sigma(m^2+i\epsilon)]=\gamma(m^2).$
	
	Defining the width
	\begin{equation}
		\Gamma\equiv \frac{g^2 Z_R}{16\pi m}\sqrt{1-\frac{4\mu^2}{m^2}}\,,
		\label{width}
	\end{equation}
	we have $Z_R\gamma(m^2)= m \Gamma$ for $m>2\mu$, and we can recast~\eqref{dress-propag-g^2} up to order $\mathcal{O}(g^2)$ as 
	\begin{equation}
		\bar{G}_{\phi,a}(z)\simeq \frac{iaZ_R}{z-m^2+ia m \Gamma}\simeq \frac{iaZ_R}{z-\left(m-\frac{i}{2}a\Gamma\right)^2}\,.
		\label{resummed-narrow-width}
	\end{equation}
	The last expression has a complex pole at
	\begin{equation}
		z =  m^2-i a m \Gamma\,,\label{generic-complex-pole}
	\end{equation}
	 whose imaginary part's sign is equal to that of $-a$ since $\Gamma>0.$ The corresponding complex mass is $m-ia\Gamma/2$. The computation of the dressed propagator in eq.~\eqref{resummed-narrow-width} is valid around $z\simeq m^2,$ this means that the result is true in the so-called narrow-width approximation, i.e. for $\Gamma\ll m.$ 
	
	The absorptive part of the propagator~\eqref{resummed-narrow-width} for real momentum square reads
	\begin{equation}
		{\rm Im}\left[i\bar{G}_{\phi,a}(s,b\epsilon)\right]\simeq  \frac{ab\epsilon Z_R+m\Gamma Z_R}{(s-m^2)^2+(b\epsilon+a  m \Gamma)^2}\,,
		\label{imag-narrow-width}
	\end{equation}
	and the two contributions~\eqref{A-function} and~\eqref{B-function} in the narrow-width approximation become
	\begin{eqnarray}
		A_a(s,b\epsilon)\!\!\!&\simeq&\!\!\!  \frac{ab\epsilon Z_R}{(s-m^2)^2+(b\epsilon +a m \Gamma)^2}\,,\\[1.5mm]
		\label{A-func-narrow}
		B_a(s,b\epsilon)\!\!\!&\simeq&\!\!\!  \frac{m\Gamma Z_R}{(s-m^2)^2+(b\epsilon+a m \Gamma)^2}\,.
		\label{B-func-narrow}
	\end{eqnarray}

	\subsubsection{Remarks on the limits $\epsilon\rightarrow 0^+$ and $\Gamma\rightarrow 0^+$}\label{sec:remarks-limit}
	
	When ${\rm Im}[\Sigma(m^2)]\neq 0$ one usually takes the limit $\epsilon\rightarrow 0^+$ and only keeps the contribution from $B_a$, which is proportional to the multiparticle spectral density. At the same time, one is often interested in studying the time scales during which the $\phi$-particle still propagates on-shell with an approximate real mass (e.g., time scales during which an ordinary particle is still alive), that is, processes related to~$A_a.$  Strictly speaking, this physical scenario cannot be described because our QFT framework is defined in an infinite interval of time. This means that any unstable particle would have already decayed since infinite time has passed. 
		
	What is typically done to study unstable particles before their decay takes place is to work in the narrow-width approximation, make an expansion in $\Gamma/m\ll 1$ and eventually take the limit $\Gamma\rightarrow 0^+$ to get an on-shell Dirac delta as a function of the renormalized mass. However, taking the limits $\Gamma\rightarrow 0^+$ and $\epsilon\rightarrow 0^+$ naively could lead to physical and mathematical inconsistencies, in particular when $a=-1$. Let us clarify this aspect.
	
	If we take the limit $\epsilon\rightarrow 0^+$ of eq.~\eqref{imag-narrow-width} we get
	\begin{eqnarray}
		\lim\limits_{\epsilon\rightarrow 0^+}{\rm Im}\left[i\bar{G}_{\phi,a}(s,b\epsilon)\right]\simeq   \frac{m\Gamma Z_R}{(s-m^2)^2+(m \Gamma)^2}\,,
		\label{eps-limit}
	\end{eqnarray}
	which is independent of $b.$ Although the contribution from $A_a$ disappeared, one usually considers time scales over which the particle still propagates on-shell with an approximate real mass by working in the regime $\Gamma/m\ll 1$; in particular, in the limit $\Gamma\rightarrow 0^+$ one obtains
	\begin{eqnarray}
		\lim\limits_{\Gamma\rightarrow 0^+}\lim\limits_{\epsilon\rightarrow 0^+}{\rm Im}\left[i\bar{G}_{\phi,a}(s,b\epsilon)\right]\simeq   \pi Z_R \delta(s-m^2)\,.
		\label{eps-gamma-limit}
	\end{eqnarray}
	For an ordinary particle ($a=1=b$) the limit~\eqref{eps-gamma-limit} happens to match with the expression we would expect for a $\phi$-particle propagating on-shell. However, this is just a lucky coincidence, in fact the Dirac delta contribution must come from the absorptive term $A_a$, which diagrammatically accounts for the cuts of the $\phi$-lines as shown in eq.~\eqref{A-function-diagram}.
	
	The correct procedure to study regimes in which a resonance can still be approximated as a particle propagating on-shell with a real mass (e.g., regimes in which an ordinary particle is still alive) is to first take the limit $\Gamma\rightarrow 0^+$ and then $\epsilon\rightarrow 0^+.$ In this case we have
	\begin{eqnarray}
		\lim\limits_{\epsilon\rightarrow 0^+} \lim\limits_{\Gamma\rightarrow 0^+}{\rm Im}\left[i\bar{G}_{\phi,a}(s,b\epsilon)\right]\simeq \lim\limits_{\epsilon\rightarrow 0^+} \frac{ab\epsilon Z_R}{(s-m^2)^2+\epsilon^2}=   ab\pi Z_R \delta(s-m^2)\,,
		\label{gamma-eps-limit}
	\end{eqnarray}
	and the result of the limit is now given by $A_a.$ Therefore, the limits $\epsilon\rightarrow 0^+$ and $\Gamma\rightarrow 0^+$ do \textit{not} commute and the two double limits have two different  meanings. In the case of an ordinary particle the results of the two double limits~\eqref{eps-gamma-limit} and~\eqref{gamma-eps-limit} coincide by accident. The distinction becomes evident for ghost particles ($a=-1$) and may lead to ambiguities. 
	
	For instance, the double limit~\eqref{eps-gamma-limit} of the Feynman-ghost dressed propagator ($a=-1$, $b=1$) gives a positive absorptive contribution $+\pi Z_R \delta(s-m^2)$, unlike its tree-level counterpart that has the opposite sign. This fact is sometimes used to claim that quantum corrections convert the Feynman-ghost into a particle which propagates positive energies backward in time. However, from our discussion above we know that this cannot be possible because the on-shell contribution in the narrow-width approximation must be studied by taking the double limit~\eqref{gamma-eps-limit}. Indeed, if we do so, we obtain  $-\pi Z_R \delta(s-m^2)$ which is consistent with the causal propagation of the Feynman-ghost propagator, that is, with positive energies propagating forward in time. Therefore,~\eqref{gamma-eps-limit} is the correct limit to take in order to describe time scales during which a Feynman-ghost particle behaves as a narrow resonance propagating on-shell with real mass $m$.
	
	On the other hand, for an anti-Feynman ghost the limit~\eqref{gamma-eps-limit} gives $+\pi Z_R\delta(s-m^2)$, which is compatible with positive energies propagating backward in time. Indeed, anti-Feynman-ghost particles propagate acausally, that is, they are characterized by an arrow of time opposite to that of ordinary particles~\cite{Donoghue:2019fcb,Donoghue:2019ecz}. Similarly to  the case of ordinary particles, also for an anti-Feynman ghost the two double limits~\eqref{gamma-eps-limit} and~\eqref{eps-gamma-limit} coincide by accident.

	In summary, if we want to study a narrow resonance for time scales during which the particle still propagates on-shell with an approximate real mass $m$ (e.g., during which an ordinary particle is still alive), we must first take the limit $\Gamma\rightarrow 0^+$ and then $\epsilon\rightarrow 0^+.$ The opposite order would give the wrong physical result. It is also worth to mention that the more rigorous approach is to work with a QFT formulated in a finite interval of time $\Delta t<\infty$ and consider time scales $\Delta t\lesssim 1/\Gamma$. In this case, it would be possible to consistently study physical situations such that $A_a$ is non-zero and $B_a$ vanishes without the need to make any forced approximation~\cite{Anselmi:2020lfx}. In sec.~\ref{sec:finite-time-QFT}, we will further elaborate on the importance to work with finite-time QFTs to capture the true physical features of ghost resonances.

	\subsection{Poles locations and Riemann sheets}
	
	We now want to determine the poles locations of the dressed propagator~\eqref{resummed-propag}. First, we will consider the standard case with no ghost $(a=1)$ and explain several details that are usually omitted in QFT textbooks (books where some details are provided are~\cite{Eden:1966dnq,Brown:1992db,Coleman:2018mew}). Second, we will analyse the scenario with ghost ($a=-1$).

	\subsubsection{Equation for the poles to order $\mathcal{O}(g^2)$} 
	
	To understand whether the pole~\eqref{generic-complex-pole} appears in the first or second Riemann sheet, we have to study when the denominator of the dressed propagator~\eqref{resummed-propag} vanishes, i.e. whether
	\begin{equation}
	\quad \quad \,\,z-m^2+a\Sigma(z)=0
	\label{denom=0}
	\end{equation}
	or
	\begin{equation}
	\qquad z-m^2+a\Sigma^{II}(z)=0\,,
		\label{denom=0-second}
	\end{equation}
	where the self-energy in the second sheet $\Sigma^{II}(z)$ was introduced in eq.~\eqref{self-second}.
	
	Since we are working in perturbation theory, up to order $\mathcal{O}(g^2)$ and after imposing the renormalization conditions~\eqref{renorm-cond}, the algebraic equation~\eqref{denom=0} can be written in the following form:
	\begin{equation}
		z-m^2=-ia \frac{g^2 Z_R}{16\pi} \sqrt{1-\frac{4\mu^2}{z}}\theta\left({\rm Re}[z]-2\mu\right)\,,
		\label{algebrain-eq-generic}
	\end{equation}
	where the square root in first sheet gets a positive (negative) sign if continued onto the real axis from above (below). As an ansatz for the complex pole solution we consider
	\begin{equation}
		z=m^2e^{i\varphi}\,,
		\label{sol-exp-phase}
	\end{equation}
	where the phase $\varphi$ must be determined by solving~\eqref{algebrain-eq-generic} up to order $\mathcal{O}(g^2)$; in particular, $\varphi$ will depend linearly on $\Gamma/m$.
	
	The algebraic equation~\eqref{algebrain-eq-generic} with the ansatz~\eqref{sol-exp-phase} is defined in the first sheet for $\varphi\in [0,2\pi).$ On the other hand, if we consider values of $\varphi$ in the ranges $[-2\pi,0)$ or $[2\pi,4\pi)$ we end up in the second sheet. Since the square root is two-sheeted we can move from one sheet to the other every $2\pi$-rotation. This means that we can look for solutions in both sheets by solving~\eqref{algebrain-eq-generic} and finding in which range of values the phase $\varphi$ falls. For example, a solution of~\eqref{denom=0} with $\varphi=\tilde{\varphi}\in [-2\pi,0)$ corresponds to a pole in the second sheet and is equivalent to the solution~\eqref{sol-exp-phase} of~\eqref{denom=0-second} with $\varphi=\tilde{\varphi}+2\pi \in [0,2\pi).$

	\subsubsection{Standard case without ghost}\label{sec:2-no-ghost}
	
	We first consider the case of an ordinary $\phi$-field, i.e. we solve eq.~\eqref{algebrain-eq-generic} with $a=1:$
	\begin{equation}
			z-m^2=-i \frac{g^2 Z_R}{16\pi} \sqrt{1-\frac{4\mu^2}{z}}\theta\left({\rm Re}[z]-2\mu\right)\,.
		\label{algebrain-eq-standard}
	\end{equation}
	If $m^2\leq 4\mu^2$, eq.~\eqref{algebrain-eq-standard} is trivially solved: the propagator has a pole on the real axis at $z=m^2$ and its analytic structure is shown in Fig.~\ref{fig1.1}. If $m^2>4\mu^2$, we already know that eq.~\eqref{algebrain-eq-standard} admits the complex solution $z=m^2-i m \Gamma,$ where the width $\Gamma$  was defined up to order $\mathcal{O}(g^2)$ in eq.~\eqref{width}. We now want to show that the complex pole appears in the second Riemann sheet.

	\begin{itemize}
		
		\item \textbf{First Riemann sheet.} If the complex pole were located in the fourth quadrant of the first sheet, then it could be reached from the first quadrant by going anti-clock-wise around the branch point, as shown in Fig.~\ref{fig1.2}. In this case we would choose $\varphi=2\pi-\Gamma/m$ with $\Gamma/m\ll 1,$ thus $m^2e^{i(2\pi-\Gamma/m)}\simeq m^2-im\Gamma$ and the left-hand side of eq.~\eqref{algebrain-eq-standard} would be
		\begin{equation}
			z-m^2=-im\Gamma + \mathcal{O}(g^4) \,.
			\label{lhs-first-standard}
		\end{equation}
		Whereas, the right-hand side of~\eqref{algebrain-eq-standard} acquires the phase factor $e^{-i\varphi/2}=e^{-i(2\pi-\Gamma/m)/2}\simeq -1-i\Gamma/(2m)$ due to the square root and reads
		\begin{equation}
			-i\frac{g^2 Z_R}{16\pi} \sqrt{1-\frac{4\mu^2}{z}}= i\frac{g^2 Z_R}{16\pi} \sqrt{1-\frac{4\mu^2}{m^2}}+\mathcal{O}(g^4)=im\Gamma+\mathcal{O}(g^4)\,.
			\label{rhs-first-standard}
		\end{equation}
		Its sign is opposite to~\eqref{lhs-first-standard}: this means that the algebraic equation~\eqref{algebrain-eq-standard} is not solved by $z=m^2e^{i(2\pi-\Gamma/m)}$ with $\Gamma/m\ll 1,$ that is, there is no pole in the first sheet. 
		
		
		\begin{figure}[t!]
			\centering
			\subfloat[Subfigure 1 list of figures text][]{
				\includegraphics[scale=0.33]{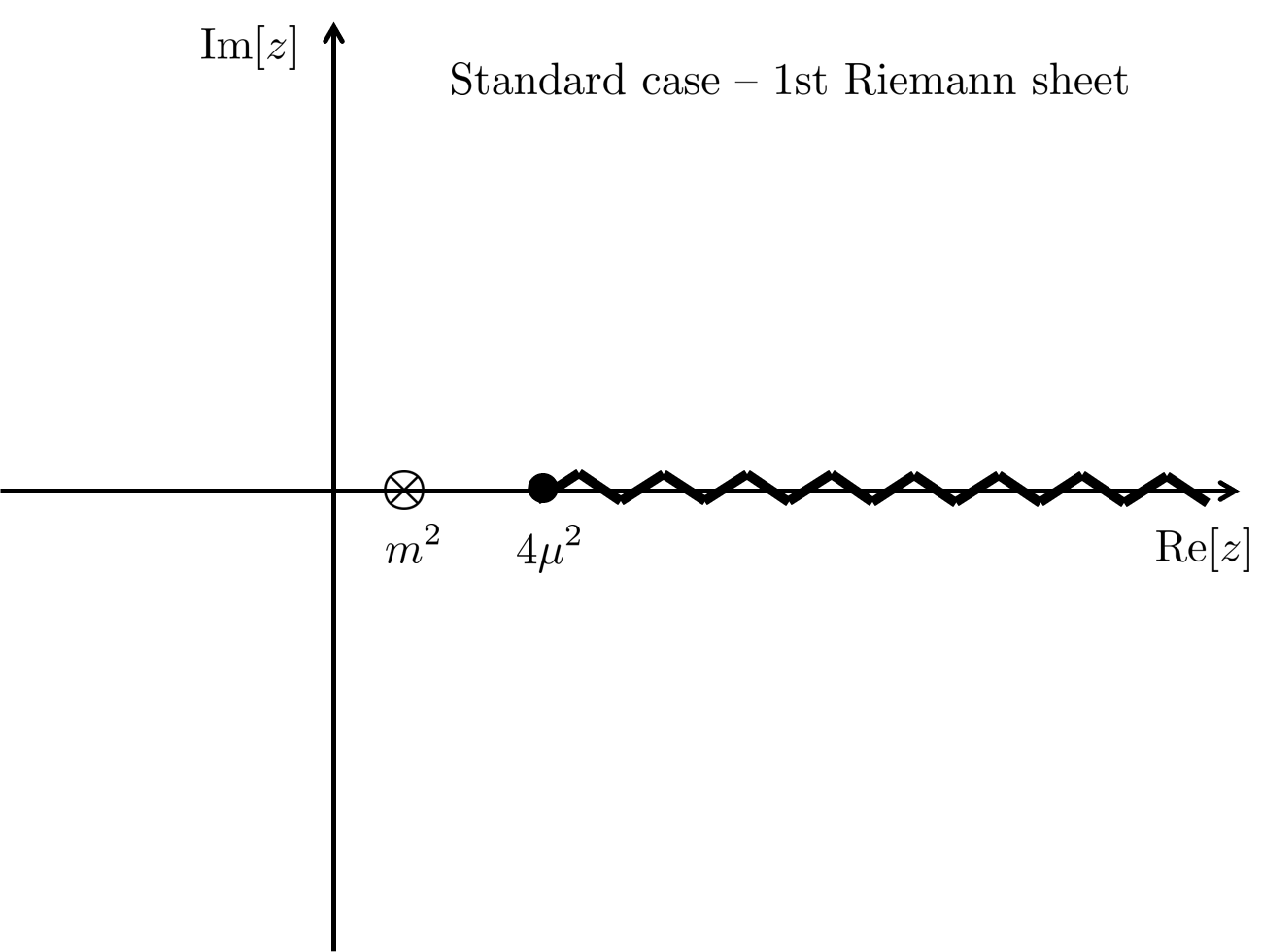}\label{fig1.1}}\qquad\,
			\subfloat[Subfigure 2 list of figures text][]{
				\includegraphics[scale=0.33]{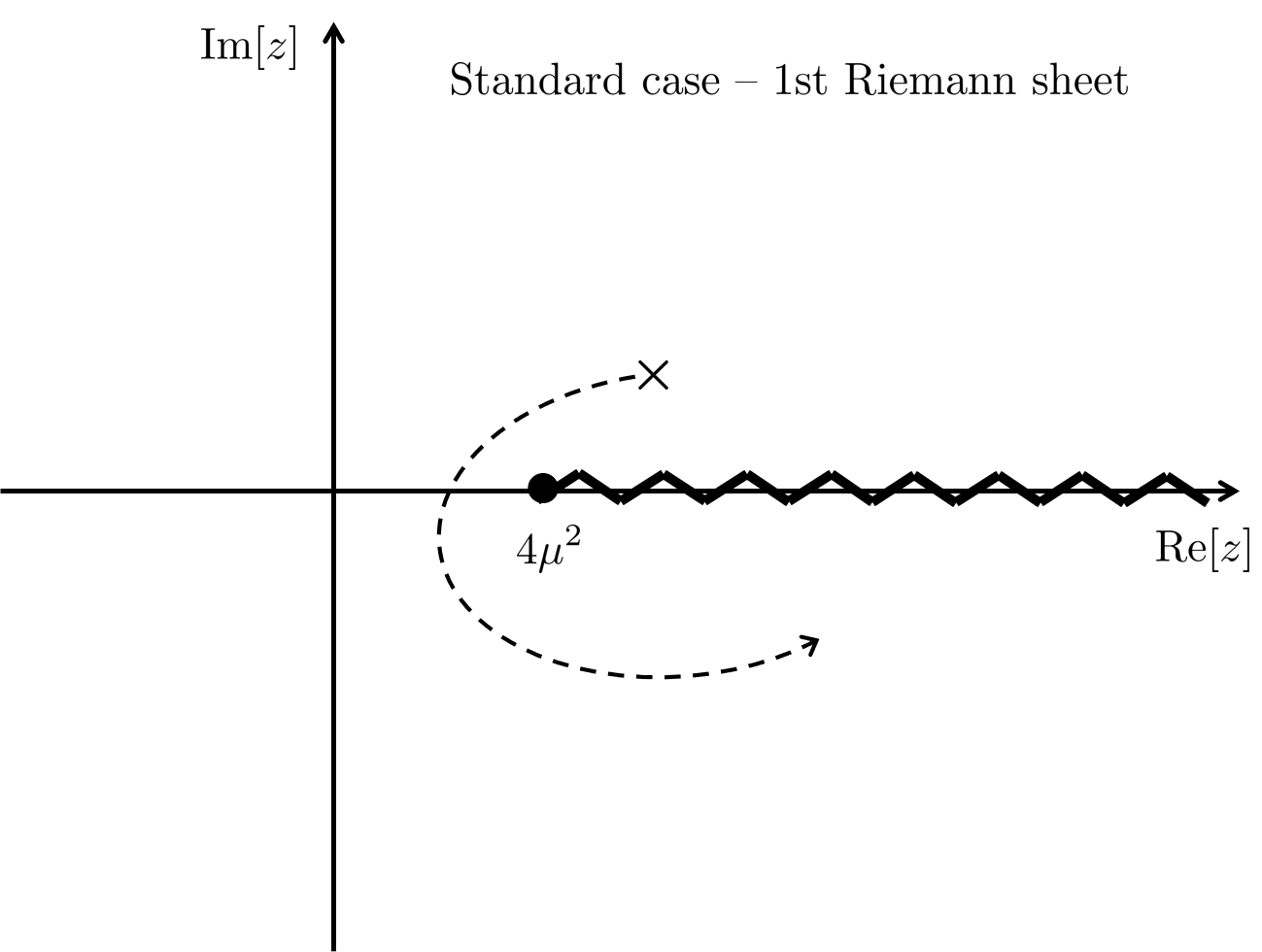}\label{fig1.2}}\\
			\subfloat[Subfigure 2 list of figures text][]{
				\includegraphics[scale=0.33]{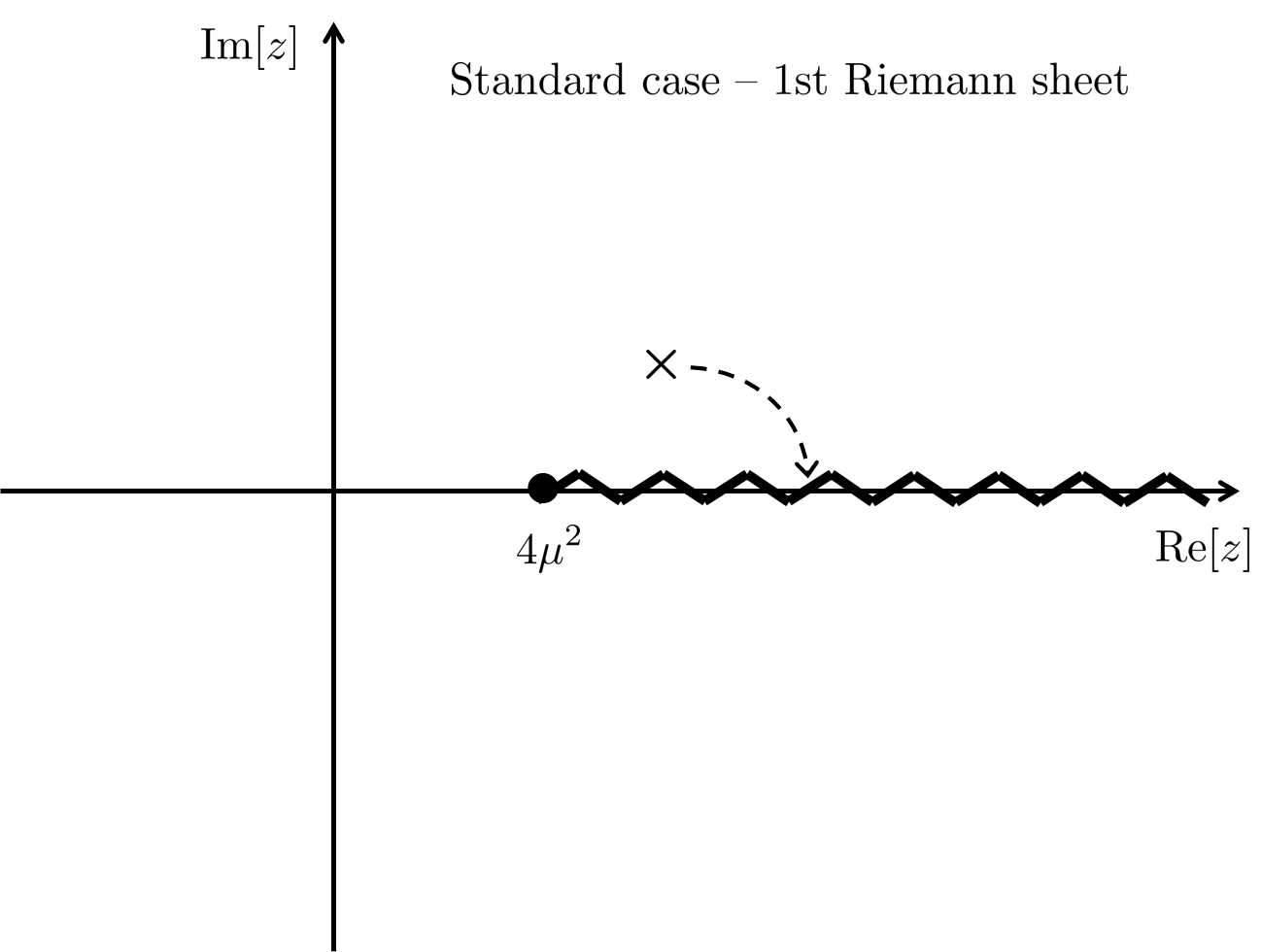}\label{fig1.3}}\qquad\,
			\subfloat[Subfigure 2 list of figures text][]{
				\includegraphics[scale=0.33]{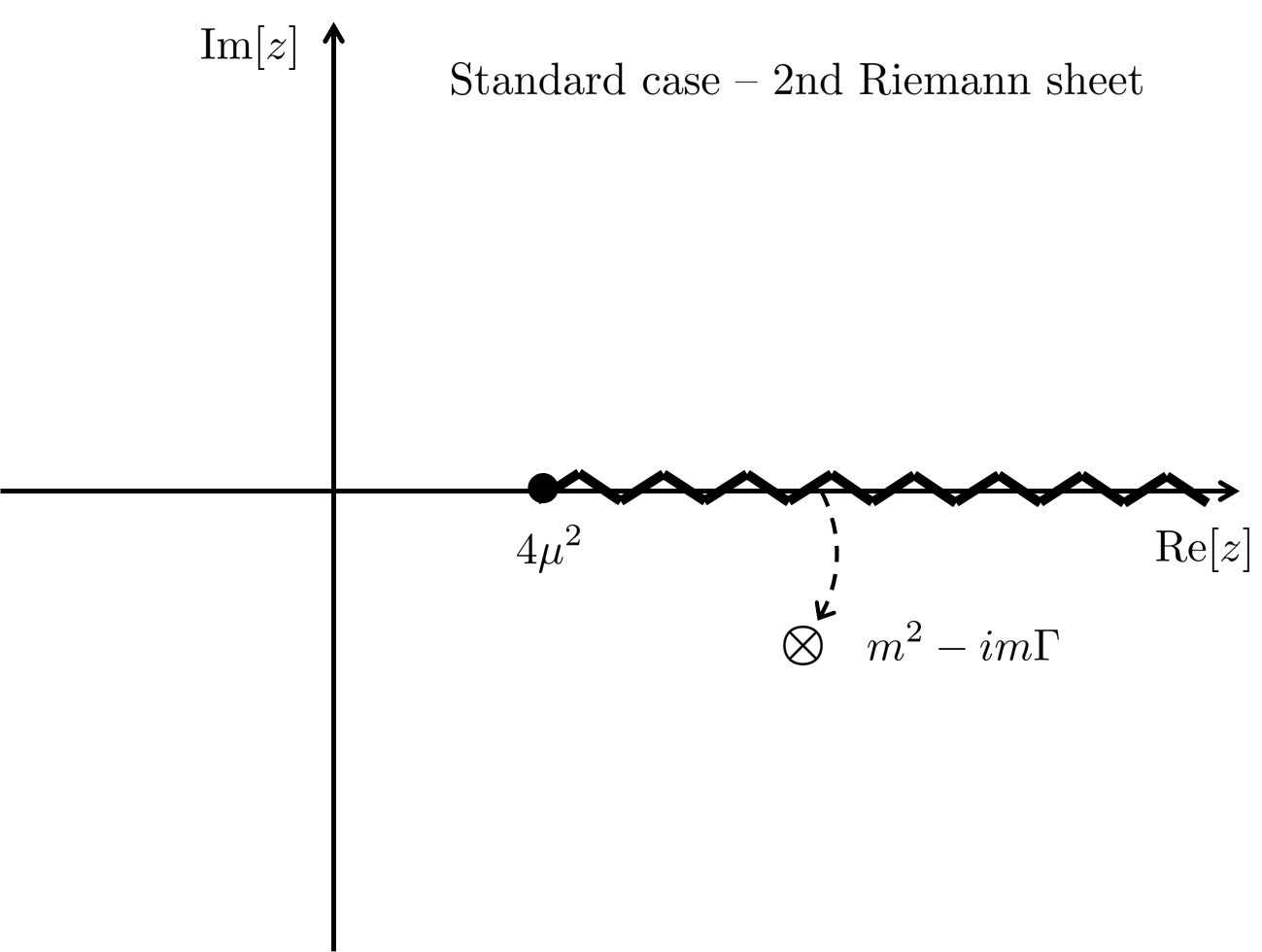}\label{fig1.4}}
			\protect\caption{Analytic structure of the ordinary $\phi$-propagator with $a=1$. (a)~First Riemann sheet showing an isolated pole on the real axis in addition to the branch cut when  $m^2< 4\mu^2$. For $m^2=4\mu^2$ the pole sits on the branch point. (b)~Anti-clock-wise path around the branch point to go below the real axis in the fourth quadrant while remaining in the first Riemann sheet. The pole $z=m^2-im\Gamma$ is not found in the first sheet.  (c)-(d)~Clock-wise path to go below the real axis passing through the branch cut and ending up in the fourth quadrant of the second Riemann sheet. In this case a pole at $z=m^2-im\Gamma$ with $m^2>4\mu^2$ is found. The encircled cross is a pole.}
			\label{fig1}
		\end{figure}
		
		
		\item \textbf{Second Riemann sheet.} The other possibility to reach the region below the real axis is to proceed clock-wise. This means that we pass through the branch cut and end up in the fourth quadrant of the second Riemann sheet, as shown in Figs.~\ref{fig1.3} and~\ref{fig1.4}. In this case the phase is  $\varphi=-\Gamma/m$ with $\Gamma/m\ll 1,$ which still gives $z=m^2 e^{-i\Gamma/m}\simeq m^2-im\Gamma$ but from the square root we now get the phase factor $e^{i\Gamma/(2m)}\simeq 1+i\Gamma/(2m).$ Therefore, the left-hand side of~\eqref{algebrain-eq-standard} is still given by~\eqref{lhs-first-standard}, while the right-hand side  now reads
		\begin{equation}
			-i\frac{g^2 Z_R}{16\pi} \sqrt{1-\frac{4\mu^2}{z}}=-im\Gamma+\mathcal{O}(g^4)\,.
			\label{rhs-second-standard}
		\end{equation}
		The two sides now have the same signs, that is, eq.~\eqref{algebrain-eq-standard} is solved up to order $\mathcal{O}(g^2)$ and the complex pole $m^2-im\Gamma$ appears in the second Riemann sheet, as shown in Fig.~\ref{fig1.4}.
		
		Furthermore, the reflection property  $\Sigma^*(z)=\Sigma(z^*)$ implies that the complex conjugate mass square $m^2+im\Gamma$ must also be a solution of~\eqref{algebrain-eq-standard} and thus a pole of the propagator. This pole appears in the first quadrant of the second Riemann sheet and can be reached from the fourth quadrant of the second sheet by going clock-wise around the branch point. Indeed, if we take $\varphi=-(2\pi-\Gamma/m)$ we have $z=m^2 e^{-i(2\pi-\Gamma/m)}\simeq m^2+im\Gamma$ and from the square root we get the phase factor $e^{i(2\pi-\Gamma/m)/2}\simeq -1+i\Gamma/(2m).$  Thus, the left- and right-hand sides of~\eqref{algebrain-eq-standard} up to order $\mathcal{O}(g^2)$ are equal and the equation is solved.

	\end{itemize}

\begin{figure}[t!]
		\centering
		\subfloat[Subfigure 1 list of figures text][]{
			\includegraphics[scale=0.33]{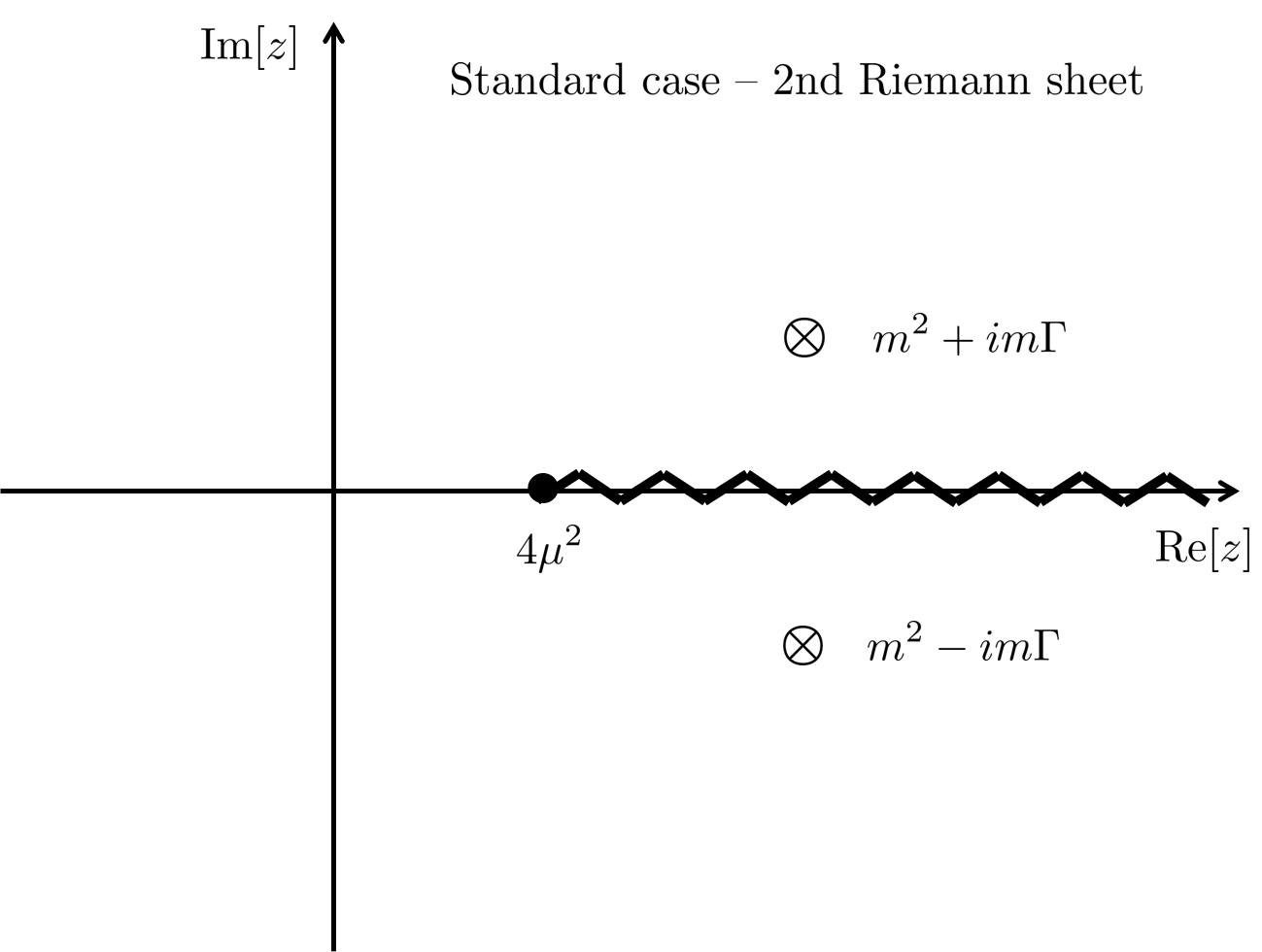}\label{fig2.1}}\qquad\,
		\subfloat[Subfigure 2 list of figures text][]{
			\includegraphics[scale=0.33]{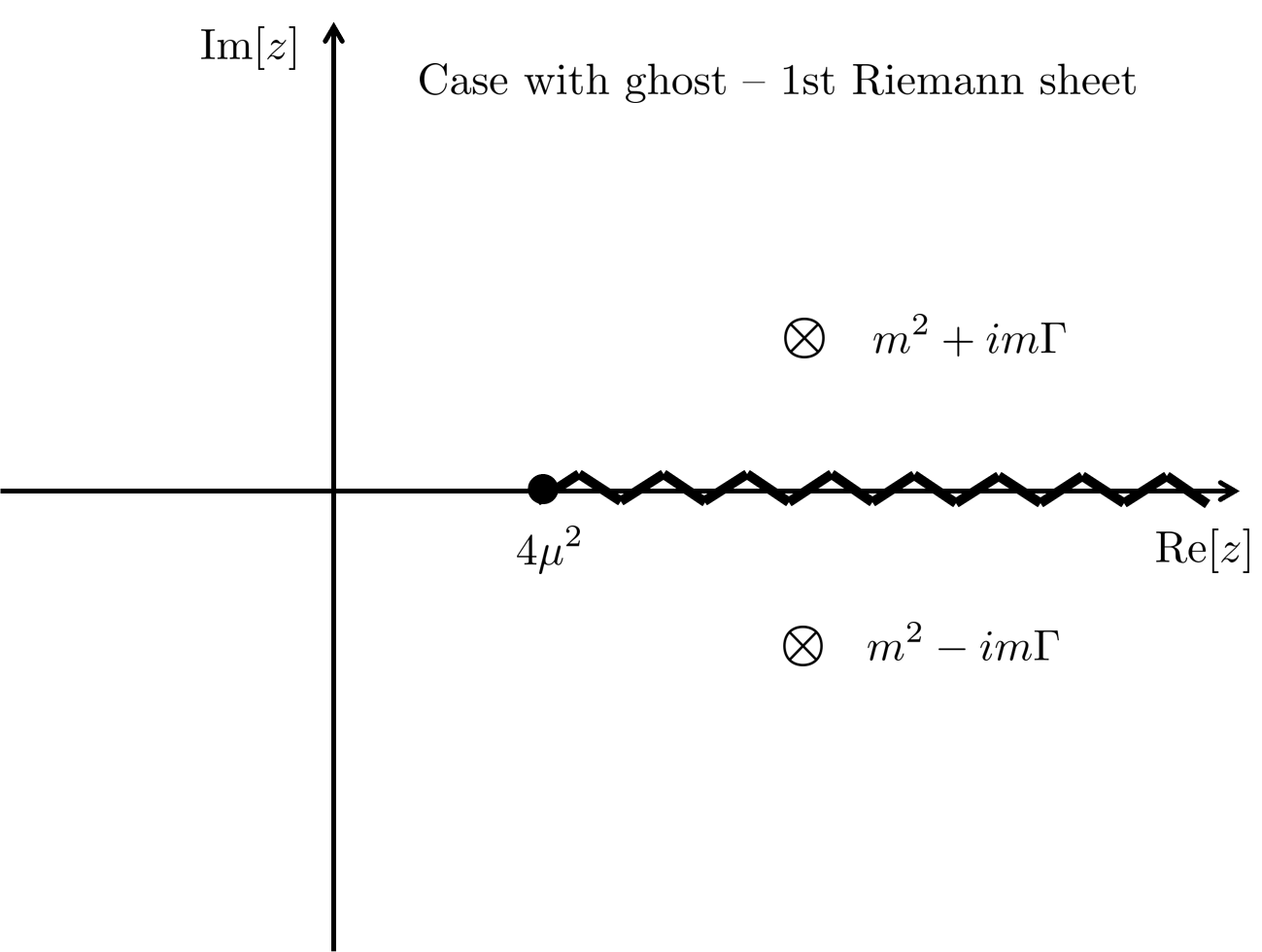}\label{fig2.2}}
		\protect\caption{(a) Analytic structure of the ordinary propagator in the second Riemann sheet: the pair of complex conjugate poles $m^2\pm m\Gamma$ appears when $m^2> 4\mu^2.$ (b) Analytic structure of the ghost propagator in the first Riemann sheet: the pair of complex conjugate poles $m^2\pm m\Gamma$ appears when $m^2> 4\mu^2$. Encircled crosses are poles.}
		\label{fig2}
	\end{figure}

	We have shown that for $m^2>4\mu^2$ an ordinary $\phi$-propagator has a pair of complex conjugate poles $m^2\pm im\Gamma$ in its second Riemann sheet, as shown in Fig.~\ref{fig2.1}.  It is also worth to mention that an ordinary propagator can in general have additional poles in the higher Riemann sheets, but in the model under investigation we have the square root to order $\mathcal{O}(g^2)$, which is two-sheeted. This means that if we go clock-wise through the branch cut from the first quadrant of the second sheet we end up again in the first sheet where there is no pole, that is, the ansatz $\varphi=-(2\pi+\Gamma/m)$ does not solve eq.~\eqref{algebrain-eq-standard}.
	
	Another question that could be asked is why we do not see the complex conjugate pole from the narrow-width approximation performed above, that is, why we only see the pole $z=m^2-im\Gamma$ in eq.~\eqref{resummed-narrow-width} when $a=+1$. First of all, we must emphasize again that the expression~\eqref{resummed-narrow-width} does \textit{not} have a pole in the first Riemann sheet. In fact, what is a pole from the point of view of the second sheet is a bump in the spectral density from the point of view of the first sheet. This can be understood from the K\"allén–Lehmann spectral representation of the propagator above the multiparticle threshold and in the first sheet:
	\begin{equation}
		\bar{G}_{\phi,+}(z)=\int_{4\mu^2}^\infty {\rm d}\sigma \rho(\sigma)\frac{i}{z-\sigma}\,,
		\label{spectral-standard-first}
	\end{equation}
	where $\rho(\sigma)$ is the spectral density that is positive and non-zero for $\sigma> 4\mu^2$, and satisfies the relation
	\begin{equation}
		\rho(s)=\frac{1}{\pi}{\rm Im}\left[i\bar{G}_{\phi,+}(s+i\epsilon)\right]\simeq \frac{1}{\pi}\frac{m\Gamma Z_R}{(s-m^2)^2+(m \Gamma)^2}=\frac{Z_R}{2\pi i}\left[\frac{1}{s-m^2-im\Gamma}-\frac{1}{s-m^2+im\Gamma}\right]\,,
		\label{spectral-peack-standard}
	\end{equation}
	where we used~\eqref{imag-narrow-width} with $\epsilon\rightarrow 0^+.$ From eq.~\eqref{spectral-peack-standard} we understand that the spectral density is highly peaked in the narrow-width approximation, i.e. for $s\simeq m^2$ and  $\Gamma/m\ll 1$.
	
	
	An expression with the pole-like structure like~\eqref{resummed-narrow-width} can be derived from the spectral representation~\eqref{spectral-standard-first} by making the following approximation. Since $\rho(s)$ in eq.~\eqref{spectral-peack-standard} is peaked around $s\simeq m^2,$ we can approximately extend the integration range in~\eqref{spectral-standard-first} up to $-\infty$ and then use the residue theorem to obtain
	\begin{equation}
		\bar{G}_{\phi,+}(z)\simeq \int_{-\infty}^\infty {\rm d}\sigma \frac{Z_R}{2\pi i}\left[\frac{1}{\sigma-m^2-im\Gamma}-\frac{1}{\sigma-m^2+im\Gamma}\right]\frac{i}{z-\sigma}=\frac{i Z_R}{z-m^2+im\Gamma}\,,
		\label{spectral-standard-first-approx}
	\end{equation}
	where we have taken ${\rm Im}[z]>0$ since the narrow-width approximation was defined through a Taylor expansion in a region around $z\simeq m^2$, just above the real axis in the first Riemann sheet and just below the real axis in the second sheet. The pole-like structure in eq.~\eqref{spectral-standard-first-approx} is just an artefact of the narrow-width approximation.
	
	On the other hand, from the point of view of the second Riemann sheet, it so happens that the contribution of the spectral density approximately cancels that of the isolated complex conjugate pole $m^2+im\Gamma$, and this is why we only see the effect of the pole $m^2-im\Gamma.$ This explicit cancellation will be shown in the next section after defining the propagator in the second sheet in a model-independent way.

	\subsubsection{Case with ghost}\label{sec:2-feynman}
	
	We now consider the case of a ghost $\phi$-field, i.e. we solve eq.~\eqref{algebrain-eq-generic} with $a=-1$:
	\begin{equation}
		z-m^2=i \frac{g^2 Z_R}{16\pi} \sqrt{1-\frac{4\mu^2}{z}}\theta\left({\rm Re}[z]-2\mu\right)\,.
		\label{algebrain-eq-feyn-gh}
	\end{equation}
	If $m^2\leq 4\mu^2$, eq.~\eqref{algebrain-eq-feyn-gh} is trivially solved: the propagator has a pole on the real axis at $z=m^2$ and its analytic structure is the same as the ordinary case shown in Fig.~\ref{fig1.1}. If $m^2>4\mu^2$, we know that eq.~\eqref{algebrain-eq-feyn-gh} admits the complex solution $z=m^2+i m \Gamma.$

	\begin{itemize}
		
		\item \textbf{First Riemann sheet.} To show that the complex pole $m^2+i m \Gamma$ appears in the first quadrant of the first sheet we have to verify that the ansatz~\eqref{sol-exp-phase} with $\varphi=\Gamma/m$ is a solution. Since $e^{i\Gamma/m}\simeq 1+i\Gamma/m$ and $e^{-i\Gamma/(2m)}\simeq 1-i\Gamma/(2m),$ the left- and right-hand sides of~\eqref{algebrain-eq-feyn-gh} read
		\begin{equation}
			z-m^2=im\Gamma + \mathcal{O}(g^4)
			\label{lhs-first-feyn-gh}
		\end{equation}
		and
		\begin{equation}
			i\frac{g^2 Z_R}{16\pi} \sqrt{1-\frac{4\mu^2}{z}}=im\Gamma + \mathcal{O}(g^4)\,,
			\label{rhs-first-feyn-gh}
		\end{equation}
		respectively. The two expressions have the same signs, that is,~\eqref{algebrain-eq-feyn-gh} is solved to  order $\mathcal{O}(g^2)$ and the complex pole $m^2+im\Gamma$ appears in the first Riemann sheet.
		
		Furthermore, the reflection property  $\Sigma^*(z)=\Sigma(z^*)$ implies that the complex conjugate mass square $m^2-im\Gamma$ must also be a solution of~\eqref{algebrain-eq-feyn-gh}. This pole appears in the fourth quadrant of the first Riemann sheet and can be reached from the first quadrant by going anti-clock-wise around the branch point. Indeed, if we take $\varphi=(2\pi-\Gamma/m)$ we have $z=m^2 e^{i(2\pi-\Gamma/m)}\simeq m^2-im\Gamma$ and from the square root we get the phase factor $e^{-i(2\pi-\Gamma/m)/2}\simeq -1-i\Gamma/(2m).$  Thus, the left- and right-hand sides of~\eqref{algebrain-eq-feyn-gh} are equal and the equation is solved.

		\item  \textbf{Second Riemann sheet.} It can be shown that the ghost propagator does not have any complex pole in the second Riemann sheet. If the complex pole $m^2-im\Gamma$ were located in the fourth quadrant of the second sheet, then it could be reached from the first quadrant of the first sheet by going clock-wise and passing through the branch cut, i.e. with $\varphi=-\Gamma/m$ such that $m^2e^{-i\Gamma/m}\simeq m^2-im\Gamma$ and $e^{i\Gamma/(2m)}\simeq 1+i\Gamma/(2m).$ However, this ansatz does not solve~\eqref{algebrain-eq-feyn-gh}.
		The same can be shown for the pole $m^2+im\Gamma$.
		
	\end{itemize}

	We have proven that for $m^2>4\mu^2$ a ghost propagator has a pair of complex conjugate poles $m^2\pm im\Gamma$ in its first Riemann sheet, as shown in Fig.~\ref{fig2.2}.  
	As before, another question that could be asked is why in the narrow-width approximation we only see the pole $m^2+im\Gamma$ in eq.~\eqref{resummed-narrow-width} when $a=-1$. It so happens that there is an approximate cancellation between the spectral density contribution and the pole $m^2-im\Gamma.$ This will be explicitly shown in the next section after defining the spectral representation for the ghost propagator.

	\subsubsection{Summary of poles locations}
	
	We can summarize our analysis of poles locations as follows. From eq.~\eqref{generic-complex-pole} and the property $\Sigma^*(z)=\Sigma(z^*),$ we know that propagator has the pair of complex conjugate poles $m^2\pm i m\Gamma$, where the expression of the width $\Gamma$ up to order $\mathcal{O}(g^2)$ is given in~\eqref{width}. We have 
	\begin{eqnarray}
	(m^2\pm im\Gamma)-m^2+a{\rm Re}\left[\Sigma(m^2\pm im\Gamma)\right]= \pm i Z^{-1}_R m\Gamma+\mathcal{O}(g^4)\,,
	\end{eqnarray}
	which is valid in both first and second Riemann sheets.
	
	Moreover, from the analytic structure of the self-energy we know that
	\begin{eqnarray}
	\text{First sheet:}&& {\rm Im}\left[\Sigma(m^2\pm im\Gamma)\right]=\pm Z_R^{-1} m\Gamma +\mathcal{O}(g^4)\,,\\[1.5mm]
	\text{Second sheet:}&& {\rm Im}\left[\Sigma^{II}(m^2\pm im\Gamma)\right]=\mp Z_R^{-1} m\Gamma+\mathcal{O}(g^4)\,.
	\end{eqnarray}
	Therefore, substituting $m^2\pm i m\Gamma$ in~\eqref{denom=0} and~\eqref{denom=0-second}, up to the order $\mathcal{O}(g^2)$ we obtain:
	\begin{eqnarray}
	\text{First sheet:}&& \pm 1= \mp a \,,\label{1st-sheet-eq}\\[1.5mm]
	\text{Second sheet:}&& \pm 1= \pm a \,.\label{2nd-sheet-eq}
	\end{eqnarray}

	If $a=1$,~\eqref{1st-sheet-eq} is not solved, while~\eqref{2nd-sheet-eq} has a solution: this means that an ordinary dressed propagator has no complex poles in the first Riemann sheet but it has a pair of complex conjugate poles in the second sheet. If $a=-1$,~\eqref{1st-sheet-eq} has a solution, while~\eqref{2nd-sheet-eq} is not solved: the pair of complex conjugate poles appears in the first sheet and no pole is present in the second sheet. This conclusion is independent of the prescription that is used to define the propagator on the real axis.

	\section{Spectral representations}\label{sec:general-analysis}
	
	The results of the previous section were obtained by working with the QFT model in eq.~\eqref{lagrangian}. The aim of this section is to describe the same results in a model-independent way using general features of the propagator such as the spectral representation. We will explicitly find the analytic continuation of the propagator into the second Riemann sheet and define spectral representations in both first and second sheets.
	
	To simplify the notation, we now call the propagator simply $\bar{G}_a(z)$. We do not refer to any particular Lagrangian, but we still assume that there is some $\phi$-field that can acquire a non-vanishing width through interactions with ordinary fields.

	\subsection{Standard case without ghost}
	
	The ordinary propagator $\bar{G}_+(z)$ is an analytic function of $z\in \mathbb{C}$ except for isolated poles and branch cuts on the real axis. These analytic features are captured by the K\"allén–Lehmann spectral representation which provides a very useful expression for the propagator in the first Riemann sheet. For real masses above the threshold for multiparticle production we have 
	\begin{equation}
		\bar{G}_{+}(z)=\int_{M_{\rm th}^2}^\infty{\rm d}\sigma \rho(\sigma)\frac{i}{z-\sigma}\,,
		\label{spectral-standard}
	\end{equation}
	where $\rho(s)=\frac{1}{\pi}{\rm Im}[i\bar{G}_{+}(s+i\epsilon)]$ is the spectral density which is positive and non-zero above the threshold, i.e. for ${\rm Re}[z]=s>M^2_{\rm th}.$ The mass square $M^2_{\rm th}$ is the lowest multiparticle threshold which coincides with $4\mu^2$ in the field theory model considered in the previous section. The propagator satisfies the following reflection property:
	\begin{equation}
		\bar{G}^*_{+}(z)=-\bar{G}_{+}(z^*)\,.
		\label{reflection-property-standard}
	\end{equation}

	The spectral representation~\eqref{spectral-standard} clearly shows that an ordinary propagator does not have any pole in the first Riemann sheet for real masses above the multiparticle threshold, and the only type of non-analyticity is a branch cut on the real axis with branch point at $z=M^2_{\rm th}$. We now want to find the explicit expression of the propagator in the second sheet and prove that it has a pair of complex conjugate poles there.
	
	\paragraph{Continuation into the second Riemann sheet.} Let us consider the inverse of the propagator and define the corresponding spectral density $\omega(s)$:
	\begin{equation}
	\omega(s)\equiv \frac{1}{\pi}{\rm Im}\left[\frac{1}{i}\bar{G}_+^{-1}(s+i\epsilon)\right]\,.
		\label{inverse-propag-standard}
	\end{equation}
	Using $\bar{G}^{-1}_+=\bar{G}^*_+/|\bar{G}_+|^2$ and~\eqref{reflection-property-standard} we can write
	\begin{equation}
		\omega(s) =\frac{1}{\pi}\frac{1}{|\bar{G}_+(s+i\epsilon)|^2}{\rm Im}\left[i\bar{G}_+(s-i\epsilon) \right]=-\frac{1}{|\bar{G}_+(s+i\epsilon)|^2}\rho(s)\leq 0\,.
		\label{omega-rho-standard}
	\end{equation}
	The imaginary part of $\bar{G}_+^{-1}$ is also related to the discontinuity across the branch cut, 
	\begin{equation}
		\frac{1}{i}\left[\bar{G}^{-1}_+(s+i\epsilon)-\bar{G}^{-1}_+(s-i\epsilon)\right]=2\pi i\, \omega(s)\,.
		\label{discont-standard}
	\end{equation}

	We now have all the ingredients to define the propagator in the second Riemann sheet. Starting in the upper half of the first sheet, we can analytically continue downward into the second sheet by defining  the function $G^{II\,-1}_+$ that just below the real axis in the second sheet is equal to the function $G^{-1}_+$  just above the real axis in the first sheet. In formula we have~\cite{Brown:1992db}
	\begin{equation}
		\begin{aligned}
		\frac{1}{i}\bar{G}_+^{II\,-1}(s-i\epsilon)\equiv & \,\,\frac{1}{i}\bar{G}_+^{-1}(s+i\epsilon) \\[1mm]
		=&\,\, \frac{1}{i}\bar{G}_+^{-1}(s-i\epsilon) +i2\pi\omega(s)\,,
		\label{contin-to-second-standard}
		\end{aligned}
	\end{equation}
	where in the second step we used~\eqref{discont-standard}. Thus, the propagator in the second Riemann sheet and just below the real axis reads
	\begin{eqnarray}
		\bar{G}^{II}_+(s-i\epsilon) =\frac{-i}{\frac{1}{i}\bar{G}^{-1}_+(s-i\epsilon)+i2\pi\omega(s)}=\frac{\bar{G}_+(s-i\epsilon)}{1-\bar{G}_+(s-i\epsilon)2\pi \omega(s)}\,.
		\label{propag-second-standard}
	\end{eqnarray}
	Moreover, if we analytic continue $\omega(s)$ into the complex plane we can define the propagator in the second sheet as a function of $z\in \mathbb{C}:$ 
	\begin{eqnarray}
		\bar{G}^{II}_+(z) =\frac{-i}{\frac{1}{i}\bar{G}^{-1}_+(z)-i2\pi\omega(z)}=\frac{\bar{G}_+(z)}{1+\bar{G}_+(z)2\pi \omega(z
			)}\,,
		\label{propag-second-standard-z-complex}
	\end{eqnarray}
	where $\omega(s\pm i\epsilon)=\pm \omega(s).$

	\paragraph{Complex conjugate poles in the second sheet.} Although $\bar{G}^{-1}_+(z)$ has no zeros above the multiparticle threshold, the function $\frac{1}{i}\bar{G}^ {-1}_+(z)-i2\pi\omega(z)$ has complex zeros.
	
	Let us work in the narrow-width approximation and expand in Taylor series around $z\simeq m^2$ and just below the real axis. Taking ${\rm Re}[\bar{G}^{-1}_+(m^2)]=0$ as definition of the physical real mass, defining
	\begin{eqnarray}
		Z_R^{-1}\equiv -\left.{\rm Re}\left[\frac{1}{i}\frac{{\rm d}\bar{G}_+^{-1}(z)}{{\rm d}z}\right]\right|_{z=m^2}>0\,,
		\label{Z-constant-standard}
	\end{eqnarray}
	and using ${\rm Im}\big[\frac{1}{i}\bar{G}_+^{-1}(m^2-i\epsilon)\big]=-\omega(m^2)$, eq.~\eqref{propag-second-standard-z-complex} can be approximately written as
	\begin{eqnarray}
		\bar{G}^{II}_+(z) \simeq \frac{iZ_R}{z-m^2-i\pi Z_R\,\omega(m^2)}\,.
		\label{propag-second-standard-3}
	\end{eqnarray}
	Note that we have neglected the first derivative of $\omega(z)$ evaluated in the region around $z\simeq m^2$ because it contributes to higher orders.
	
	From eq.~\eqref{omega-rho-standard} we know that the spectral function $\omega(s)$ is negative. Thus, it is convenient to define the following positive quantity that will have the meaning of a width:
	\begin{eqnarray}
		\Gamma\equiv -\frac{\pi Z_R\,\omega(m^2)}{m}>0\,, 
		\label{width-standard}
	\end{eqnarray}
	through which we can recast the propagator just below the real axis in the second sheet as
	\begin{eqnarray}
		\bar{G}^{II}_+(z) \simeq \frac{iZ_R}{z-m^2+im\Gamma}\,.
		\label{propag-second-standard-4}
	\end{eqnarray}
	This expression shows a pole $z=m^2-im\Gamma$ in the second Riemann sheet.
	
	Furthermore, from the reflection property $\bar{G}^*_+(z)=-\bar{G}_+(z^*)$ it follows that if $z$ is a zero of the inverse propagator in the second sheet, then $z^*$ will also be a zero and will solve the algebraic equation $\frac{1}{i}\bar{G}^{-1}_+(z^*)+i2\pi \omega(z^*)=0.$ In fact, working in the narrow-width approximation it can be easily found that $m^2+im\Gamma$ is also a pole. 
	
	In summary, we have shown that an ordinary propagator ($a=+1$) has a pair of complex conjugate poles in the second Riemann sheet. This result is non-perturbative and model-independent, and obviously in agreement with the conclusions reached in sec.~\ref{sec:2-no-ghost} using perturbation theory and resumming the self-energies.
	
	\paragraph{Spectral representation in the second sheet.} We now want to derive a spectral representation for an ordinary propagator in the second sheet and above the multiparticle threshold. From the discussion above, we expect the presence of two isolated complex conjugate poles $m^2\pm im\Gamma$. Moreover,  the relation
	\begin{eqnarray}
		\frac{1}{\pi}{\rm Im}\left[\frac{1}{i}\bar{G}_+^{II\,-1}(s-i\epsilon)\right]=\frac{1}{\pi}{\rm Im}\left[\frac{1}{i}\bar{G}_+^{-1}	(s+i\epsilon)\right]=\omega(s)
	\end{eqnarray}
	suggests that we can also associate the spectral density $\rho(s)$ to the propagator in the second sheet, that is, we have $\frac{1}{\pi}{\rm Im}\left[i\bar{G}_+^{II}(s-i\epsilon)\right]=\rho(s)$.

	
	\begin{figure}[t!]
		\centering
		\includegraphics[scale=0.37]{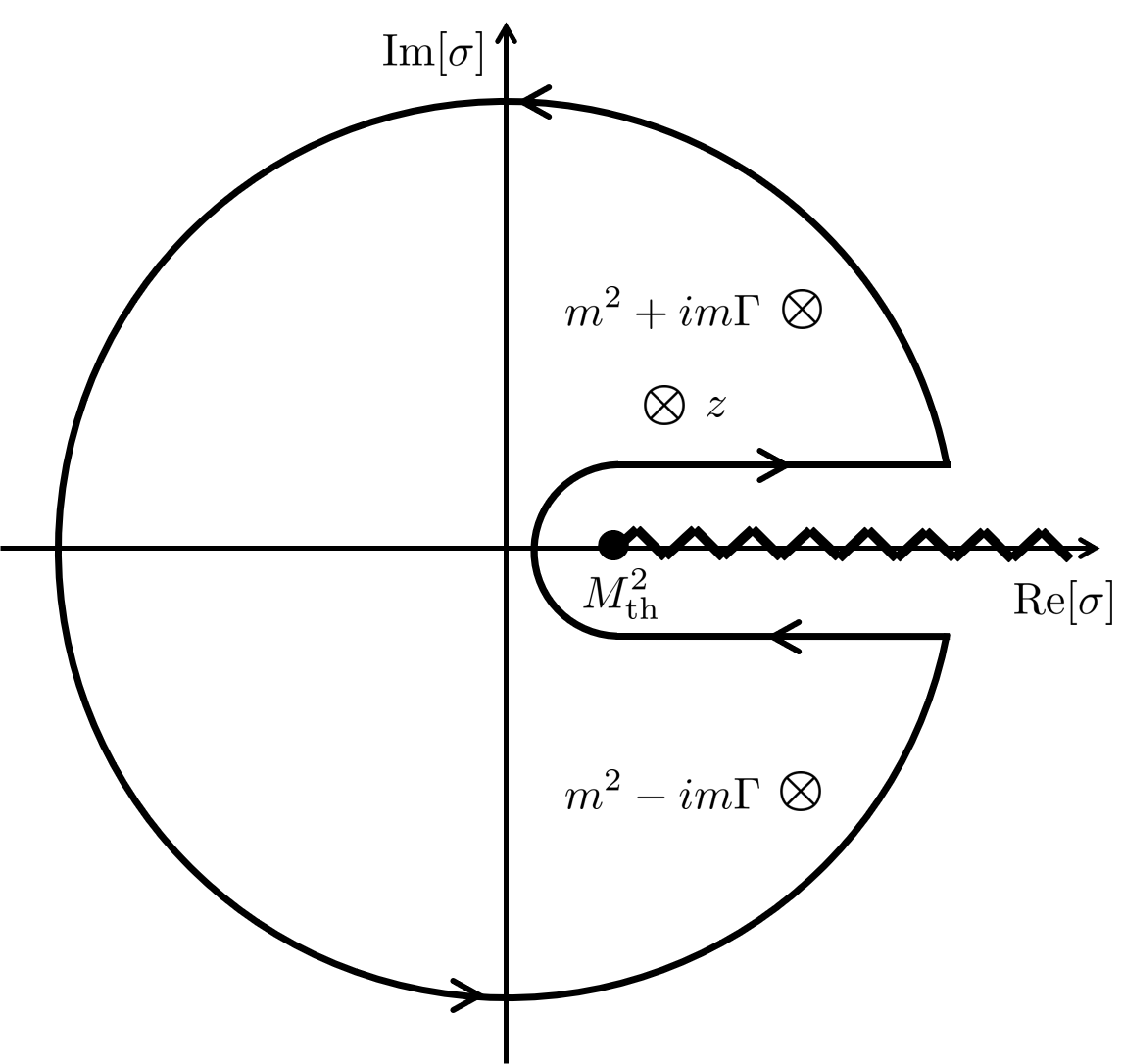}
		\protect\caption{Integration contour $\mathcal{C}$ to derive the spectral representations above the multiparticle threshold  for a standard propagator in the second Riemann sheet, i.e. eq.~\eqref{spectral-standard-second}, and  for the ghost propagator in the first sheet, i.e. eq.~\eqref{spectral-feyn-gh}. Encircled crosses are poles. The pole $z$ satisfies ${\rm Re}[z]>M_{\rm th}^2$ and can lie either in the upper or lower half plane.}
		\label{fig4}
	\end{figure}
	
	
	Indeed, computing the following integral with the residue theorem,
	\begin{eqnarray}
		I_+(z)=\frac{1}{2\pi i}\int_{\mathcal{C}}{\rm d}\sigma \frac{\bar{G}_+^{II}(\sigma)}{\sigma-z}
		\label{cont-int-standard}
	\end{eqnarray}
	where $\mathcal{C}$ is the contour in Fig.~\ref{fig4}, and using the fact that the integral is also equal to discontinuity across the branch cut,
	\begin{equation}
		I_+(z)=\frac{1}{2\pi i} \lim\limits_{\varepsilon\rightarrow 0^+}\int_{M^2_{\rm th}}^{\infty}{\rm d}\sigma\left[\bar{G}^{II}_+(\sigma+i\varepsilon)-\bar{G}^{II}_+(\sigma-i\varepsilon)\right]\frac{1}{\sigma-z}=\int_{M^2_{\rm th}}^\infty{\rm d}\sigma\frac{-i\rho(\sigma)}{z-\sigma}\,,
		\label{int-disc-standard}
	\end{equation}
	we obtain the spectral representation
	\begin{equation}
		\bar{G}^{II}_{+}(z)=\frac{iZ}{z-(m^2+ i m\Gamma)}+\frac{iZ^*}{z-(m^2- i m\Gamma)}+\int_{M_{\rm th}^2}^\infty{\rm d}\sigma \rho(\sigma)\frac{-i}{z-\sigma}\,,
		\label{spectral-standard-second}
	\end{equation}
	where $m^2>M^2_{\rm th}$, $Z$ and $Z^*$ are the residues of the two complex conjugate poles. In general, the residues have both real and imaginary parts, i.e. $Z=Z_R+i Z_I$, but in the narrow-width approximation the imaginary part can be neglected, thus we approximately have $Z,Z^*\simeq Z_R.$ 
	
	It is worth to note that the multiparticle integral contributions to the spectral representations in the first~\eqref{spectral-standard} and second~\eqref{spectral-standard-second} Riemann sheets differ by an overall opposite sign.

	We are now able to answer a question we asked at the end of sec.~\ref{sec:2-no-ghost} about why only the complex pole $m^2-im\Gamma$ dominates in the expression of the propagator in the second sheet when we work in the narrow-width approximation in the field theory model~\eqref{lagrangian} with $a=1$. We can use a reasoning similar to the one that led us to eq.~\eqref{spectral-standard-first-approx}. Since for a narrow resonance the spectral density $\rho(s)$ is highly peaked around $s\simeq m^2$ as shown in eq.~\eqref{spectral-peack-standard}, we can approximately extend the integration range from $M^2_{\rm th}$ to $-\infty$ and write
	\begin{eqnarray}
		\bar{G}^{II}_{+}(z)\!\!\!&\simeq &\!\!\!\frac{iZ_R}{z-(m^2+ i m\Gamma)}+\frac{iZ_R}{z-(m^2- i m\Gamma)}\nonumber \\[1.5mm]
		&&\!\!\!+\int_{-\infty}^\infty{\rm d}\sigma \frac{Z_R}{2\pi i}\left[\frac{1}{\sigma-m^2-im\Gamma}-\frac{1}{\sigma-m^2+im\Gamma}\right]\frac{-i}{z-\sigma}\,,
		\label{second-standard-approx}
	\end{eqnarray}
	where we have also used $Z,Z^*\simeq Z_R$. The narrow-width approximation was defined through a Taylor expansion in a neighbourhood of $z\simeq m^2$ which includes a region of the first quadrant in the first Riemann sheet and a region of the fourth quadrant in the second sheet. Thus, in eq.~\eqref{second-standard-approx} we have to take ${\rm Im}[z]<0$ and,  applying the residue theorem, we obtain
	\begin{eqnarray}
		\int_{-\infty}^\infty{\rm d}\sigma \frac{Z_R}{2\pi i}\left[\frac{1}{\sigma-m^2-im\Gamma}-\frac{1}{\sigma-m^2+im\Gamma}\right]\frac{-i}{z-\sigma}=\frac{-iZ_R}{z-(m^2+im\Gamma)}\,,
		\label{integral-standard}
	\end{eqnarray}
	which cancels the first term in~\eqref{second-standard-approx}, so that we only obtain a contribution from the pole $m^2-im\Gamma$ as in eq.~\eqref{propag-second-standard-4}. However, we must emphasize that this cancellation is only valid as an approximation for a narrow resonance. In fact, the integral contribution in eq.~\eqref{second-standard-approx} does not have any pole and acquires a pole-like structure only in the narrow-width approximation.

	\subsection{Case with ghost}
	
	Let us now consider the ghost case ($a=-1$). In sec.~\ref{sec:2-feynman}, we saw that in the model~\eqref{lagrangian} the ghost propagator had a pair of complex conjugate poles in its first Riemann sheet, above the multiparticle threshold. This suggests that $\bar{G}_-(z)$ is an analytic function of $z\in \mathbb{C}$ except for isolated complex conjugate poles and branch cuts. Furthermore, the analysis in sec.~\ref{sec:2-feynman} also shows that above the multiparticle threshold the analytic structure of the ghost propagator is the same for both Feynman and anti-Feynman prescriptions. Indeed, we found that the multiparticle absorptive contribution $B_-(s,b\epsilon)$ was independent of $b$ (see eqs.~\eqref{B-function} and~\eqref{B-func-narrow}). Therefore, the following discussion will apply to both Feynman and anti-Feynman ghost propagators.

	\paragraph{Spectral representation in the first sheet.}  
	
	Consider
	\begin{eqnarray}
		I_-(z)=\frac{1}{2\pi i}\int_{\mathcal{C}}{\rm d}\sigma \frac{\bar{G}_-(\sigma)}{\sigma-z}\,,
		\label{cont-int-feyn-gh}
	\end{eqnarray}
	where $\mathcal{C}$ is the contour in Fig.~\ref{fig4}. Combining the computation of the integral through the residue theorem and the fact that 
	\begin{equation}
		I_-(z)=\frac{1}{2\pi i} \lim\limits_{\varepsilon\rightarrow 0^+}\int_{M^2_{\rm th}}^{\infty}{\rm d}\sigma\left[\bar{G}_-(\sigma+i\varepsilon)-\bar{G}_-(\sigma-i\varepsilon)\right]\frac{1}{\sigma-z}=\int_{M^2_{\rm th}}^\infty{\rm d}\sigma\frac{i\rho(\sigma)}{z-\sigma}\,,
		\label{int-disc-feyn-gh}
	\end{equation}
	we obtain the following spectral representation of the ghost propagator in the first sheet:
	\begin{equation}
		\bar{G}_{-}(z)=\frac{-iZ}{z-(m^2+ i m\Gamma)}+\frac{-iZ^*}{z-(m^2- i m\Gamma)}+\int_{M_{\rm th}^2}^\infty{\rm d}\sigma \rho(\sigma)\frac{i}{z-\sigma}\,,
		\label{spectral-feyn-gh}
	\end{equation}
	where the spectral density is given by $\rho(s)=\frac{1}{\pi}{\rm Im}\big[\bar{G}_-(s+i\epsilon)\big]$, $Z$ and $Z^*$ are the residues at the two complex conjugate poles $m^2\pm i m\Gamma$, and $m^2>M^2_{\rm th}$.
	
	From eq.~\eqref{spectral-feyn-gh} it is also clear that the propagator satisfies the following reflection property:
	\begin{equation}
		\bar{G}^*_{-}(z)=-\bar{G}_{-}(z^*)\,.
		\label{reflection-property-gh}
	\end{equation}

	We can now answer a question we asked at the end of sec.~\ref{sec:2-feynman} about why only the complex pole $m^2+im\Gamma$ survives in the expression of the propagator in the first sheet when we work in the narrow-width approximation in the field theory model in~\eqref{lagrangian} with $a=-1$. Indeed, if the width is narrow we can write $Z,Z^*\simeq Z_R$ and
	\begin{eqnarray}
		\int_{M^2_{\rm th}}^\infty{\rm d}\sigma \frac{i\rho(\sigma)}{z-\sigma}&\simeq& 	\int_{-\infty}^\infty{\rm d}\sigma \frac{Z_R}{2\pi i}\left[\frac{1}{\sigma-m^2-im\Gamma}-\frac{1}{\sigma-m^2+im\Gamma}\right]\frac{i}{z-\sigma}\nonumber \\[1.5mm]
		&=&\frac{iZ_R}{z-(m^2-im\Gamma)}\,,
		\label{integral-feyn-gh}
	\end{eqnarray}
	where in the last integral we have taken ${\rm Im}[z]>0$ since the narrow-width approximation was defined through a Taylor expansion in a region around $z\simeq m^2$ just above the real axis in the first Riemann sheet and just below the real axis in the second sheet. The expression~\eqref{integral-feyn-gh}  approximately cancels the second term in~\eqref{spectral-feyn-gh}. Thus, for a ghost propagator ($a=-1$) only the pole $m^2+im\Gamma$ dominates in the narrow-width approximation in agreement with eqs.~\eqref{resummed-narrow-width} and~\eqref{generic-complex-pole}.

	\paragraph{Continuation into the second Riemann sheet.} To find the expression of the ghost propagator in the second sheet we can consider the same procedure that we applied to the case of the ordinary propagator, that is, we can follow the steps from~\eqref{inverse-propag-standard} to~\eqref{contin-to-second-standard}. By doing so, we get
	\begin{eqnarray}
		\bar{G}^{II}_-(z)=\frac{-i}{\frac{1}{i}\bar{G}^{-1}_-(z)-i2\pi \omega(z)} =\frac{\bar{G}_-(z)}{1+\bar{G}_-(z)2\pi \omega(z)}\,,
		\label{propag-second-gh-complex}
	\end{eqnarray}
	where $\omega(s\pm i \epsilon)=\pm \omega(s).$

	\paragraph{Spectral representation in the second sheet.} Working in the narrow-width approximation, the expression~\eqref{propag-second-gh-complex} would acquire the same pole-like structure as that of the ghost propagator in the first sheet. However, we should remark that $\bar{G}^{II}_-(z)$ does not have any pole, but the pole-like structure in the second sheet is just an artefact of the spectral density contribution. Indeed, if we take the limit $z\rightarrow m^2 \pm im\Gamma$ with $\Gamma/m\ll 1,$ the propagator in the first sheet $\bar{G}_-(z)$ diverges and we get
	\begin{equation}
	\lim\limits_{z\rightarrow m^2\pm im\Gamma} G^{II}_-(z)\sim \pm \frac{1}{2\pi\omega(m^2)}\,,
	\end{equation}	
	which does not have poles since $\omega(m^2)$ is strictly negative for $m^2>M^2_{\rm th}$.
	
	Therefore, the spectral representation of the ghost propagator in the second Riemann sheet can be written as
	\begin{eqnarray}
		\bar{G}^{II}_-(z) = \int_{M^2_{\rm th}}^\infty {\rm d}\sigma \frac{-i\rho(\sigma)}{z-\sigma}\,.
		\label{spectral-feyn-gh-second}
	\end{eqnarray}
	If we work in the narrow-width approximation and use~\eqref{integral-feyn-gh} we get the desired pole-like structure $-iZ_R/(z-m^2-im\Gamma)$. For a ghost propagator, what is a pole from the point of view of the first sheet is a bump in the spectral density from the point of view of the second sheet. This behavior is opposite, in terms of Riemann sheets, to the ordinary case without ghost.
	
	\subsection{Summary of spectral representations}\label{sec:spectral}
	
	In the previous subsections we found the spectral representations above the multiparticle threshold ($m^2>M^2_{\rm th}$) for both ordinary and ghost propagators in the first and second Riemann sheets. We now want to summarise their expressions in both sheets through two compact formulas.

	\begin{itemize}
		
		\item \textbf{First Riemann sheet.} The spectral representation of the propagator $\bar{G}_a(z)$ in the first sheet and above the multiparticle threshold is
		\begin{equation}
			\bar{G}_{a}(z)=\frac{a-1}{2}\left[\frac{iZ}{z-(m^2+ i m\Gamma)}+\frac{iZ^*}{z-(m^2- i m\Gamma)}\right]+\int_{M_{\rm th}^2}^\infty{\rm d}\sigma \rho(\sigma)\frac{i}{z-\sigma}\,.
			\label{compact-first}
		\end{equation}

		\item \textbf{Second Riemann sheet.} The spectral representation of the propagator $\bar{G}^{II}_a(z)$ in the second sheet and above the multiparticle threshold is
		\begin{equation}
			\bar{G}^{II}_{a}(z)=\frac{a+1}{2}\left[\frac{iZ}{z-(m^2+ i m\Gamma)}+\frac{iZ^*}{z-(m^2- i m\Gamma)}\right]+\int_{M_{\rm th}^2}^\infty{\rm d}\sigma \rho(\sigma)\frac{-i}{z-\sigma}\,.
			\label{compact-second}
		\end{equation}

	\end{itemize}
	We can indeed check that for $a=+1$ we recover~\eqref{spectral-standard} and~\eqref{spectral-standard-second}, while for $a=-1$  we obtain~\eqref{spectral-feyn-gh} and~\eqref{spectral-feyn-gh-second}. It is worth to note that there is some kind of \textit{duality} between the propagators in the first and second sheets, indeed we formally have
	\begin{equation}
	\bar{G}_a(z)=-\bar{G}_{-a}^{II}(z)\,.
	\end{equation}
	For instance, this means that the ghost propagator in the second sheet is dual to minus the ordinary propagator in the first sheet.

	\section{Discussion}\label{sec:discuss}
	
	We now discuss the implications of the results obtained in the previous section and, at the same time, clarify some aspects that are sometimes overlooked.

	\subsection{(Un)stable resonances} \label{sec:stable-resonance}

	The first Riemann sheet is usually called the \textit{physical sheet} because its isolated poles are related to one-particle asymptotic states. Below the multiparticle threshold ($m^2<M^{2}_{\rm th}$), for both $a=\pm 1$ we have an isolated pole at $z=m^2$, which corresponds to a stable particle. If we cross the multiparticle threshold ($m^2>M^{2}_{\rm th}$) and a non-vanishing width is produced, then different features can arise depending on the values of $a$. If $a=+1$ the stable pole disappears from the first sheet and a pair of complex conjugate poles appears in the second sheet, the ordinary particle is unstable and decays, thus disappearing from the set of asymptotic states~\cite{Veltman:1963th}. If $a=-1$, the stable pole splits into a pair of complex conjugate poles that still live in the first sheet. Does this mean that a ghost resonance still belongs to the set of asymptotic states and thus cannot decay? Let us now address this question. 
	
	\paragraph{Sum rule.} Using the spectral representation for the commutator of two fields and imposing the canonical commutation relations one can derive the following sum rule for the propagator in the first Riemann sheet~\cite{Nakanishi:1972pt,Kubo:2024ysu}:
	\begin{equation}
		a=(a-1)\,Z_R + C\,,
		\label{a-b-Z-relation}
	\end{equation}
	where $Z+Z^*=2{\rm Re}[Z]=2Z_R>0$ and we have defined $C\equiv \int_{M^2_{\rm th}}^\infty{\rm d}\sigma \rho(\sigma) >0$. 
	
	The same relation can also be derived using the fact that
	\begin{equation}
		\lim\limits_{s\rightarrow \infty} \left[-i s \bar{G}_a(s\pm i\epsilon)\right]=a\,.
		\label{residue-propag}
	\end{equation}
	For example, the validity of the last equation can be verified for the dressed propagator~\eqref{resummed-propag}. Let us analyse eq.~\eqref{a-b-Z-relation}. 
	\begin{itemize}
	
	\item In the ordinary case $(a=1)$ we get $1=C$, which can be interpreted as saying that the probability of finding the physical configuration in a multiparticle state is equal to one. This means that the decay occurs and the one-particle state disappears from the set of asymptotic states. 
	
	\item For a ghost ($a=-1$) we have $1+C=2Z_R$, where $2Z_R$ can be seen as the probability of finding states that contain ghost particles with complex masses $m\pm i\Gamma/2.$ An increase in $C$ implies an increase in $Z_R$, which means that the more “decay” tries to occur, the greater the probability of finding complex-energy states becomes. This phenomenon was named \textit{anti-instability}~\cite{Kubo:2024ysu}. 
	
	\end{itemize}

	\paragraph{Remarks.} The sum rule~\eqref{a-b-Z-relation} implies that ghost particles cannot decay and suggests that ghost resonances are quantum objects very different from ordinary resonances. Let us now make some observations.
	\begin{itemize}

	\item Despite the above conclusions, it has  been claimed that ghost particles can still decay and disappear from the sets of asymptotic and intermediate states in a way consistent with unitarity~\cite{Donoghue:2019fcb}. We find this possibility difficult to realize. Indeed, the expression of the ghost propagator used in~\cite{Donoghue:2019fcb} contains no complex poles and happens to be equal to the spectral representation in the second sheet that we derived in~\eqref{spectral-feyn-gh-second}. However, this choice of spectral representation seems unjustified, since the physical Riemann sheet is the first.

	\item The presence of asymptotic states with complex masses also suggests that the Hamiltonian contains complex conjugate eigenvalues. This implies that there exist states with zero norm in the spectrum~\cite{Coleman:1969xz}. This fact is independent of the type of prescription that is used for the ghost propagator. Indeed, even if zero-norm states cannot be observed perturbatively, their presence becomes manifest non-perturbatively, e.g. after resumming the self-energies in the propagator. For example, an anti-Feynman ghost is characterized only by positive-norm states at the perturbative level, but non-perturbatively the presence of complex masses implies the existence of zero-norm states. 
	
	\item For a Feynman ghost, the impossibility of decay can also be understood as a consequence of unitarity plus the presence of negative-norm states. If quantum probabilities are conserved, then an initial negative-norm state containing a ghost particle must evolve into a final state whose norm is also negative. This means that ghosts must also be present in the final state, thus a ghost particle cannot decay and will still contribute to the set of asymptotic states. This fact has been used to claim that “physical unitarity” is violated~\cite{Kubo:2023lpz}. However, what was really meant by “physical unitarity” is that negative probabilities could appear in the final state of a scattering process, whereas the unitarity condition on the Dyson's evolution operator, i.e. the conservation of quantum probabilities, is still preserved by construction. It is worth to mention that these negative probabilities might not be problematic if the measurable physical observables are those associated to total positive probabilities. For example, this possibility was considered in~\cite{Holdom:2021hlo,Holdom:2023usn,Holdom:2024cfq}, while in Refs.~\cite{Bender:2007wu,Bender:2008gh,Mannheim:2009zj,Salvio:2015gsi,Strumia:2017dvt,Holdom:2024onr} alternative quantum probabilistic interpretations and new inner products for ghost states with zero and negative norms have been proposed.	
	
	\item The same reasoning in terms of negative norms does not apply to an anti-Feynman ghost because in this case the initial state containing a ghost particle would have a positive norm by construction. On the other hand, anti-Feynman ghost particles propagate negative energies forward in time. Therefore, if we choose the same arrow of time for both the initial and final states, then energy conservation implies that an initial state of negative energy must evolve into a final state of negative energy. This means that if an ingoing anti-Feynman ghost couples to ordinary (positive-energy) particles, the only way to have a final state with negative energy is to have ghost particles also in the asymptotic outgoing state. It would be interesting to understand whether ghost particles propagating positive energies backward in time can be coupled consistently to ordinary particles propagating positive energies forward in time. In this case, energy conservation might allow for the absence of ghosts in the final state. However, it is not yet clear how to couple particles propagating with opposite arrows of time and whether this type of interaction can be described within the operator formalism of QFT. We leave this question for a future work.

	\item Since the derivation of the sum rule~\eqref{a-b-Z-relation} does not rely on the type of prescription used for the ghost propagator, it should also apply to a fakeon ghost. As mentioned in sec.~\ref{sec:tree-unitarity}, the tree-level fakeon propagator  is an average of Feynman and anti-Feynman propagators, thus its imaginary part vanishes. Moreover, the analysis performed in~\cite{Anselmi:2020lfx} shows that the fakeon-ghost propagator has complex poles in the first Riemann sheet.  
	These facts motivate us to argue that the spectral representation of the fakeon propagator as a function of real momentum $s$, in the first sheet and above the multiparticle threshold, is given by
	\begin{equation}
		\begin{aligned}
		\bar{G}_{-,\,\text{fake}}(s)=&\,\frac{1}{2}\left[\bar{G}_-(s+i\epsilon)+\bar{G}_-(s-i\epsilon)\right] \\[1mm]
		=&\,\frac{-iZ}{s-(m^2+ i m\Gamma)}+\frac{-iZ^*}{s-(m^2- i m\Gamma)}+\text{P.V.}\left[\int_{M_{\rm th}^2}^\infty{\rm d}\sigma \frac{i\rho(\sigma)}{s-\sigma}\right]\,,
		\label{spectral-fakeon}
		\end{aligned}
	\end{equation}
	where $\text{P.V.}$ stands for the principal value. The propagator does not have an imaginary part but the spectral density $\rho(s)$ is still non-zero. It is now clear that from~\eqref{spectral-fakeon} one can derive the same sum rule~\eqref{a-b-Z-relation}, so here too the question of the existence of asymptotic states with complex masses can be raised. This aspect has not been discussed in previous works on fakeons. Therefore, we believe that further studies are needed before making definite statements on the viability of the fakeon prescription and its compatibility with the operator formalism of QFT at both the perturbative and non-perturbative levels.
	
	\end{itemize}

	\subsection{Comments on four-derivative QFTs}	
	
	In this work we only focused on two-derivative ghost fields but our conclusions also apply to theories whose Lagrangians contain fourth-order derivatives. For instance, if the kinetic term of a four-derivative theory contains a massless zero $z=0$ and a massive zero $z=m^2$, the tree-level propagator as a function of $-p^2=z\in \mathbb{C}$ will read
	\begin{equation}
	G_4(z)=\frac{im^2}{z(m^2-z)}\,.
	\end{equation}
	If the self-energy contributions are resummed, we obtain the four-derivative dressed propagator
	\begin{equation}
		\bar{G}_4(z)= \frac{i}{z-\frac{z^2}{m^2}+\Sigma(z)}\,.
		\label{four-deriv-dressed}
	\end{equation}

	Let us show that when the self-energy has a non-vanishing imaginary part, the complex poles appear in the second Riemann sheet. As a toy model we still consider the self-energy~\eqref{self-energy}, but we will also comment on a more physically interesting example related to the spin-two component of the propagator in Quadratic Gravity. 
	
	\paragraph{Narrow-width approximation.} To find the complex poles, we assume again that the width is small, thus we can Taylor expand the self-energy just above the real axis. In this four-derivative case and with the self-energy~\eqref{self-energy}, it is more convenient to work with a renormalized physical mass $\bar{m}\neq m.$ In particular, we impose the following renormalization conditions:
	\begin{equation}
	{\rm Re}\left[\Sigma(\bar{m}^2)\right]-\bar{m}^2 \left.{\rm Re}\left[\frac{{\rm d}\Sigma(z)}{{\rm d}z}\right]\right|_{z=\bar{m}^2}=0\,,\qquad \left.{\rm Re}\left[\frac{{\rm d}\Sigma(z)}{{\rm d}z}\right]\right|_{z=\bar{m}^2}=Z_R^{-1}-1\,,
	\end{equation}
	from which it follows that the renormalized physical mass square is
	\begin{equation}
	\bar{m}^2=m^2 Z_R^{-1}=m^2 +{\rm Re}[\Sigma(\bar{m}^2)]\,,
	\end{equation}
	where now $m=m(\Lambda)$ is the bare mass and depends on the renormalization scale $\Lambda.$ 
	
	Consequently, the propagator in the narrow-width approximation up to order $\mathcal{O}(g^2)$ reads
	\begin{equation}
	\bar{G}_4(z)\simeq \frac{iZ_R}{z\left(1-z/\bar{m}^2\right)+i\bar{m}\Gamma}\,,
	\label{narrow-four}
	\end{equation}
	where we have defined $\Gamma\equiv Z_R{\rm Im}[\Sigma(\bar{m}^2+i\epsilon)]/\bar{m}=Z_R\gamma(\bar{m}^2)/\bar{m}>0$ for $\bar{m}>2\mu.$ The expression~\eqref{narrow-four} is valid for $z$ above the multiparticle threshold and just above the real axis. In this expansion region the propagator has a complex pole at
	\begin{equation}
	z\simeq \bar{m}^2+i\bar{m}\Gamma\,,
	\label{pole-four}
	\end{equation}
	where higher orders $\mathcal{O}(g^4)$ are neglected.
	
	\paragraph{Complex poles in the first Riemann sheet.} It is easy to show that the pole~\eqref{pole-four} together with its complex conjugate are located in the first sheet. This follows from the fact that
	\begin{equation}
	(\bar{m}^2\pm i\bar{m}\Gamma)-\frac{(\bar{m}^2\pm i\bar{m}\Gamma)^2}{m^2}+{\rm Re}\left[\Sigma(\bar{m}^2\pm i\bar{m}\Gamma)\right]=\mp i Z_R^{-1}\bar{m}\Gamma +\mathcal{O}(g^4)
	\label{real-four}
	\end{equation}
	and
	\begin{equation}
	i{\rm Im}[\Sigma(\bar{m}^2\pm i \bar{m}\Gamma)]=\pm i Z_R^{-1}\bar{m} \Gamma+\mathcal{O}(g^4)\,.
	\label{imag-four}
	\end{equation}
	The last two expressions compensate each other in the denominator of the propagator~\eqref{four-deriv-dressed}. This cancellation occurs only in the first Riemann sheet since the signs on the right-hand side of~\eqref{imag-four} would be opposite in the second sheet. This means that the four-derivative nature of the propagator forces the complex poles to appear in the first sheet.

	\paragraph{Physically relevant example.} A more interesting example is given by the spin-two propagator in Quadratic Gravity, in particular its dressed expression that takes into account self-energy loops from light fields. Up to tensorial factors, the structure of the propagator is the same as~\eqref{four-deriv-dressed} and now the self-energy reads~\cite{Donoghue:2018izj,Donoghue:2018lmc,Donoghue:2019fcb,Donoghue:2021cza}
	\begin{equation}
	\Sigma(z)=-\kappa^2 z^2 \log\left(\frac{-z}{\Lambda^2}\right)\,,
	\label{self-quad-grav}
	\end{equation}
	where $\Lambda$ is the renormalization scale, while $\kappa^2$ is a positive constant which is proportional to the inverse of the Planck mass square and to the number of light fields running through the self-energy loops, and whose exact expression is not important for our discussion. The crucial point is that the self-energy acquires a positive (negative) imaginary part if we analytically continue onto the real axis from above (below), i.e.
	\begin{equation}
	{\rm Im}\left[\Sigma(s\pm i\epsilon)\right]=-\kappa^2 s^2 {\rm Im}\left[\log\left(\frac{-s\mp i\epsilon}{\Lambda^2}\right)\right]=\pm \pi \kappa^2 s^2\,.
	\label{imag-self-quad-grav}
	\end{equation}
	In this case the pair of complex conjugate poles is given by $\bar{m}^2\pm i\bar{m}^4\kappa^2.$ They also appear in the first Riemann sheet since equations analogous to~\eqref{real-four} and~\eqref{imag-four} are valid.

	\paragraph{Feynman vs anti-Feynman.} The presence of a branch cut allows in principle more ways to define the propagator as a function of real momentum square $s.$
	We define the Feynman and anti-Feynman four-derivative propagators as
	\begin{equation}
	\qquad \,\,\,\,\bar{G}_{4,\,{\rm F}}(s)=\frac{i}{s+i\epsilon-\frac{(s+i\epsilon)^2}{m^2}+\Sigma(s+i\epsilon)}
		\label{feynman-four}
	\end{equation}
	and
	\begin{equation}
		\bar{G}_{4,\,{\rm anti\text{-}F}}(s)=\frac{i}{s+i\epsilon-\frac{s^2}{m^2}+\Sigma(s+i\epsilon)}\,,
		\label{anti-feynman-four}
	\end{equation}
	respectively. While the Feynman propagator as a function of $s$ is defined as an analytic continuation of $\bar{G}_4(z)$ onto the real axis, the anti-Feynman propagator is defined through some non-analytic continuation. Indeed, we have
	\begin{equation}
		\bar{G}_{4,\,{\rm F}}(s)=\bar{G}_4(s+i\epsilon)\qquad \text{and}\qquad  \bar{G}_{4,\,{\rm anti\text{-}F}}(s)\neq \bar{G}_4(s+i\epsilon)\,.
		\label{anal-non-anal}
	\end{equation}
	This means that the usual Wick rotation to connect Euclidean and Minkowski results only works for the Feynman propagator. On the other hand, the anti-Feynman four-derivative propagator needs alternative contour deformations for the computation of loop integrals in perturbation theory and the evaluation of imaginary parts for the proof of unitarity~\cite{Donoghue:2019fcb}. For the latter prescription further investigation is still needed before claiming its consistency.

	The tree-level versions of the Feynman and anti-Feynman four-derivative propagators are
	\begin{equation}
		G_{4,\,{\rm F}}(s)=\frac{i}{s+i\epsilon-\frac{(s+i\epsilon)^2}{m^2}}\qquad \text{and}\qquad \bar{G}_{4,\,{\rm anti\text{-}F}}(s)=\frac{i}{s-\frac{s^2}{m^2}+i\epsilon}\,,
		\label{tree-four-propag}
	\end{equation}
	respectively. Both poles of the Feynman four-derivative propagator are shifted below the real axis with the standard Feynman shift, i.e. we have the massless pole $0-i\epsilon$ and the massive pole $m^2-i\epsilon.$ On the other hand, for the anti-Feynman four-derivative propagator the poles are $0-i\epsilon$ and $m^2+i\epsilon$. Indeed, we can also write
	\begin{equation}
	\quad\,\,\,	G_{4,\,{\rm F}}(s)=\frac{-im^2}{(s+i\epsilon)(s-m^2+i\epsilon)}=\frac{i}{s+i\epsilon}-\frac{i}{s-m^2+i\epsilon}\,,
	\end{equation}
	and
	\begin{equation}
		G_{4,\,{\rm anti\text{-}F}}(s)=\frac{-im^2}{(s+i\epsilon)(s-m^2-i\epsilon)}=\frac{i}{s+i\epsilon}-\frac{i}{s-m^2-i\epsilon}\,.
	\end{equation}
	It is now clear that in the Feynman four-derivative propagator both components are prescribed according to the Feynman shift, while in the anti-Feynman four-derivative propagator the ghost component is prescribed according to the anti-Feynman shift.  It is important to emphasize that for the anti-Feynman propagator analyticity is lost already at the perturbative level, even before resumming the self-energies. This is consistent with the discussion in the previous section.

	\paragraph{Spectral representation.} In physically relevant contexts (such as Quadratic Gravity) we expect the full propagator above the multiparticle threshold to still have the massless pole, while the real massive pole splits into a complex conjugate pair that lies in the first Riemann sheet. Therefore, the spectral representation of the four-derivative propagator above the multiparticle threshold and in the first sheet can be written as~\cite{Coleman:1969xz}
	\begin{equation}
		\bar{G}_4(z)=\frac{i}{z}+\frac{-iZ}{z-(\bar{m}^2+ i \bar{m}\Gamma)}+\frac{-iZ^*}{z-(\bar{m}^2- i \bar{m}\Gamma)}+\int_{M_{\rm th}^2}^\infty{\rm d}\sigma \frac{i\rho(\sigma)}{z-\sigma}\,.
		\label{spectral-four-deriv}
	\end{equation}
	For large real momentum square $s$ the full four-derivative propagator goes to zero as $1/s^2$. This means that if we multiply $\bar{G}_4(s\pm i\epsilon)$ by $-is$ and take the limit $s\rightarrow \infty,$ we will obtain exactly the same sum rule as~\eqref{a-b-Z-relation} with $a=-1$. Therefore, similar implications will follow about the nature of ghost resonances.\footnote{It is worth mentioning that dressed propagators with complex poles have also been considered in other contexts where the Lagrangians contain derivatives of order higher than four~\cite{Asorey:2024mkb} and/or non-polynomial form factors. The latter can already be present at the bare level~\cite{Shapiro:2015uxa,Buoninfante:2018mre,Briscese:2024tvc} or appear through non-perturbative quantum effects~\cite{Platania:2020knd,Platania:2022gtt}. Even in these cases it is important to understand in which Riemann sheet the complex poles appear and whether asymptotic states could be associated with them.} It is also important to point out that the spectral representation~\eqref{spectral-four-deriv} is the same for both Feynman and anti-Feynman four-derivative propagators, in fact their expressions are the same above the multiparticle threshold, as already noticed in the previous section.

	\subsection{On the validity of the resummation}\label{sec:inval-resum}
	
	We must emphasize that the entire discussion so far and almost all works in the literature implicitly assumed that the geometric series~\eqref{geometric-series} can be consistently resummed and give rise to the dressed propagator. However, as we briefly mentioned in sec.~\ref{sec:resummed}, when the self-energy has a non-zero imaginary part the argument of the geometric series becomes greater than one and eventually blows up in the region around $z\simeq m^2,$ thus signaling a breakdown of perturbation theory. This could mean that the resummation is not valid in that region and that the narrow-width approximation cannot be trusted since it is based on an expansion around $z\simeq m^2$.
	
	In the standard case without ghosts it is well known how to solve this problem~\cite{tHooft:1973wag}. Define the geometric series, and thus the dressed propagator, far from $z\simeq m^2,$ and then use analyticity to continue the result in the region around $z\simeq m^2.$ Then, formulate a modified perturbation theory according to which only diagrams without internal self-energies are taken into account and whose lines have dressed (and not bare) propagators attached to them. This reasoning works if the dressed propagator that defines the analytic continuation does not have any singularity in the first Riemann sheet, that is, if its denominator never vanishes. We know in fact that an ordinary dressed propagator has no pole in the first sheet, so the resummation of the geometric series can be trusted and is valid everywhere. Furthermore, since the complex pole appears in the second sheet, the particle becomes unstable and decays, so it is no longer an asymptotic state and can be projected out of the set of intermediate states in a way consistent with unitarity~\cite{Veltman:1963th}.
	
	In the case with ghost the situation is more complicated and less clear. The dressed propagator~\eqref{resummed-propag} with $a=-1$ (or the four-derivative version~\eqref{four-deriv-dressed}) has complex poles in the first Riemann sheet and very close to the real axis in the narrow-width approximation. This means that we should not be allowed to invoke analyticity to resum the geometric series~\eqref{geometric-series} around $z\simeq m^2$ when $a=-1.$ In Ref.~\cite{Anselmi:2022toe} it was claimed that the resummation could still be considered reliable for an anti-Feynman ghost. However, this statement is incorrect because the presence of complex poles in the first sheet is independent of the prescription that is used to define the propagator on the real axis. Indeed, it is important to point out that this lack of analyticity manifests itself only at the non-perturbative level and is different from the lack of analyticity that affects the anti-Feynman propagator already at the perturbative level. Therefore, the appearance of complex poles in the first sheet of a ghost propagator (or a four-derivative propagator) means that the standard analyticity structure is lost.
	
	At this point a legitimate question to ask is: are our results and previous findings in the literature on dressed ghost propagators really justified? We do not have a definite answer yet. It might actually be that the results are still correct, but then an alternative mathematical explanation is needed to justify the resummation around $z\simeq m^2$. In any case, an important point to mention is that even in the standard case without ghost the question of convergence of the geometric series does not arise if one works with a QFT formulated in a finite interval of time and considers sufficiently short time scales~\cite{Anselmi:2023wjx}. Let us comment on this aspect.

	\subsection{Need for finite-time QFT}\label{sec:finite-time-QFT}

	QFTs like the ones we have considered are formulated in an infinite interval of time, for example scattering processes are defined from an initial time $t_i=-\infty$ to a final time $t_f=\infty,$ that is, $\Delta t=t_f-t_i=\infty$. This means that, rigorously speaking, it is impossible to study scenarios in which an ordinary unstable particle is still alive. As explained in sec.~\ref{sec:remarks-limit}, what is usually done to account for the time scales during which an ordinary particle is still alive is to work in the narrow-width approximation perturbatively in $\Gamma/m\ll 1.$ However, the more physically justified and mathematically rigorous procedure would be to define the QFT in a finite interval of time $\Delta t<\infty$ and consider time scales shorter than the particle lifetime, i.e. $\Delta t\lesssim  1/\Gamma$. 
	
	This was recently done in Ref.~\cite{Anselmi:2023wjx} where both ordinary and fakeon particles were considered. It was found that the geometric series is always convergent for time scales shorter than the inverse of the width. This means that the dressed propagator for a narrow resonance can be defined without invoking analyticity as long as the time interval of interest is sufficiently short.
	
	As a future work we intend to apply this approach to Feynman and anti-Feynman ghost propagators, that is, we want to formulate the QFT in a finite interval of time and discuss features of the dressed propagator, including complex poles, asymptotic states and Breit-Wigner distribution. One of the advantages of this more rigorous analysis that we already foresee is that there will be no ambiguity in distinguishing the two absorptive contributions in the dressed propagator, as we discussed at the end of sec.~\ref{sec:remarks-limit}. This means that no confusion about (a)causal propagation and arrows of time will arise because for time scales $\Delta t\lesssim 1/\Gamma$, during which the particle still propagates on-shell with a real mass below the multiparticle threshold (e.g. during which an ordinary particle is still alive), the dominant absorptive contribution will be $A_a$.
	
	Finally, let us speculate that a finite-time QFT framework that might be useful for studying four-derivative field theories is that of Thermo Field Dynamics~\cite{Takahashi:1996zn}. In this formalism the number of degrees of freedom is doubled, and the additional ones can be ghost-like. This  has been already used to analyse ordinary unstable particles~\cite{DeFilippo:1977bk} and the vacuum structure in indefinite-metric QFTs~\cite{Rabuffo:1977va}. It would be interesting to understand whether the same formalism can be used to shed new light on ghost resonances and vacuum structure in four-derivative QFTs. We leave this exciting task for a future work.
	
    \section{Conclusions}\label{sec:concl}
	
	In this paper we analysed several properties of ghost resonances. We made a thorough study of the analytic structure of the propagator above the multiparticle threshold and determined the poles locations and the spectral representations in the first and second Riemann sheets. In particular, we clarified some aspects related to absorptive contributions, (a)casual propagation and arrows of time, absence of ghost decay, and also made critical remarks on the validity of the geometric series resummation for a ghost. Let us summarize the main results.
	\begin{itemize}
		
		\item We first worked with the QFT model~\eqref{lagrangian}: we derived the dressed propagator and determined poles locations. We started with a very detailed review of the standard case and proved that the propagator of an ordinary unstable particle does not have poles in the first Riemann sheet, but has a pair of complex conjugate poles in the second sheet. Subsequently, we considered the case of a ghost propagator and showed that the situation is opposite to the ordinary case: a pair of complex conjugate poles appears in the first sheet, while there is no pole in the second sheet. 
		
		\item We discussed physical and mathematical features of the two absorptive contributions appearing in the dressed propagator. We clarified that quantum effects cannot change the arrow of time. For example, a Feynman ghost will still propagate positive energies forward in time even if a non-vanishing width is produced through loop effects. Similarly, an anti-Feynman ghost will propagate positive energies backward in time, whether or not the loop contribution of the width is taken into account.
		
		\item Using model-independent properties of the propagator that do not necessarily rely on perturbation theory and on the model~\eqref{lagrangian}, we derived the spectral representations of the propagators in both first and second Riemann sheets. We explained that for an ordinary (ghost) propagator what is a complex pole from the point of view of the second (first) sheet, is a bump in the spectral density from the point of view of the first (second) sheet.
		
		\item We compared ordinary unstable resonances and ghost resonances. We argued that the latter cannot decay otherwise quantum probabilities would not be conserved in the case of a Feynman ghost, in agreement with the recent criticism put forward in~\cite{Kubo:2023lpz,Kubo:2024ysu}, or energy conservation would be violated in the case of an anti-Feynman ghost. We also commented on the fakeon prescription. Moreover, we considered the more interesting case of a four-derivative propagator and showed that the same conclusions apply to its ghost component.
		
	\end{itemize}

	Furthermore, we observed that almost all works in the literature (including the present one) on ghost fields and dressed propagators implicitly assumed that the resummation of the geometric series is valid in the region around the peak. However, this assumption is not justified because the presence of complex poles in the first Riemann sheet does not allow the use of analyticity to resum the series in the expansion region. Until the geometric series resummation is mathematically justified, we cannot consider the current results based on the use of the dressed propagator to be rigorously valid.
	
    We have proposed that a sensible approach to understand better mathematical and physical properties of ghost resonances is to work with a finite-time QFT. For time scales shorter than the inverse of the width, the geometric series can be resummed without the need to invoke analyticity~\cite{Anselmi:2023wjx}. This would also allow us to clearly distinguish the two distinct absorptive contributions in the dressed propagator and avoid ambiguities regarding (a)causal propagation and arrows of times. We believe that working with finite-time QFTs will help us shed new light on complex poles and asymptotic states in theories with ghosts, especially in higher-derivative QFTs. We will address these open questions in future work.


    \subsection*{Acknowledgements}
	
    I would like to thank Damiano Anselmi, John Donoghue and Bob Holdom for enlightening discussions, and John Donoghue for a long and productive email exchange. I am very grateful to Taichiro Kugo for reading an earlier draft of this work and providing critical and constructive feedback. I acknowledge financial support from the European Union’s Horizon 2020 research and innovation programme under the Marie Sklodowska-Curie Actions (grant agreement ID:~101106345-NLQG).
	


\bibliographystyle{utphys}
\bibliography{References}

\providecommand{\href}[2]{#2}\begingroup\raggedright\begin{thebibliography}{10}

\bibitem{Lee:1969fy}
T.~D. Lee and G.~C. Wick, ``{Negative Metric and the Unitarity of the S Matrix},'' \href{http://dx.doi.org/10.1016/0550-3213(69)90098-4}{{\em Nucl. Phys. B} {\bfseries 9} (1969) 209--243}.

\bibitem{Lee:1970iw}
T.~D. Lee and G.~C. Wick, ``{Finite Theory of Quantum Electrodynamics},'' \href{http://dx.doi.org/10.1103/PhysRevD.2.1033}{{\em Phys. Rev. D} {\bfseries 2} (1970) 1033--1048}.

\bibitem{Cutkosky:1969fq}
R.~E. Cutkosky, P.~V. Landshoff, D.~I. Olive, and J.~C. Polkinghorne, ``{A non-analytic S matrix},'' \href{http://dx.doi.org/10.1016/0550-3213(69)90169-2}{{\em Nucl. Phys. B} {\bfseries 12} (1969) 281--300}.

\bibitem{Coleman:1969xz}
S.~Coleman, ``{Acausality},'' in {\em {7th International School of Subnuclear Physics (Ettore Majorana): Subnuclear Phenomena}}.
\newblock 1969.

\bibitem{Nakanishi:1971jj}
N.~Nakanishi, ``{Lorentz noninvariance of the complex-ghost relativistic field theory},'' \href{http://dx.doi.org/10.1103/PhysRevD.3.811}{{\em Phys. Rev. D} {\bfseries 3} (1971) 811--814}.

\bibitem{Nakanishi:1972wx}
N.~Nakanishi, ``{Covariant formulation of the complex-ghost relativistic field theory and the lorentz noninvariance of the s matrix},'' \href{http://dx.doi.org/10.1103/PhysRevD.5.1968}{{\em Phys. Rev. D} {\bfseries 5} (1972) 1968--1975}.

\bibitem{Nakanishi:1972pt}
N.~Nakanishi, ``{Indefinite metric quantum field theory},'' \href{http://dx.doi.org/10.1143/PTPS.51.1}{{\em Prog. Theor. Phys. Suppl.} {\bfseries 51} (1972) 1--95}.

\bibitem{Boulware:1983vw}
D.~G. Boulware and D.~J. Gross, ``{LEE-WICK INDEFINITE METRIC QUANTIZATION: A FUNCTIONAL INTEGRAL APPROACH},'' \href{http://dx.doi.org/10.1016/0550-3213(84)90167-6}{{\em Nucl. Phys. B} {\bfseries 233} (1984) 1--23}.

\bibitem{Grinstein:2008bg}
B.~Grinstein, D.~O'Connell, and M.~B. Wise, ``{Causality as an emergent macroscopic phenomenon: The Lee-Wick O(N) model},'' \href{http://dx.doi.org/10.1103/PhysRevD.79.105019}{{\em Phys. Rev. D} {\bfseries 79} (2009) 105019}, \href{http://arxiv.org/abs/0805.2156}{{\ttfamily arXiv:0805.2156 [hep-th]}}.

\bibitem{Grinstein:2007mp}
B.~Grinstein, D.~O'Connell, and M.~B. Wise, ``{The Lee-Wick standard model},'' \href{http://dx.doi.org/10.1103/PhysRevD.77.025012}{{\em Phys. Rev. D} {\bfseries 77} (2008) 025012}, \href{http://arxiv.org/abs/0704.1845}{{\ttfamily arXiv:0704.1845 [hep-ph]}}.

\bibitem{Carone:2008iw}
C.~D. Carone and R.~F. Lebed, ``{A Higher-Derivative Lee-Wick Standard Model},'' \href{http://dx.doi.org/10.1088/1126-6708/2009/01/043}{{\em JHEP} {\bfseries 01} (2009) 043}, \href{http://arxiv.org/abs/0811.4150}{{\ttfamily arXiv:0811.4150 [hep-ph]}}.

\bibitem{Anselmi:2017yux}
D.~Anselmi and M.~Piva, ``{A new formulation of Lee-Wick quantum field theory},'' \href{http://dx.doi.org/10.1007/JHEP06(2017)066}{{\em JHEP} {\bfseries 06} (2017) 066},
\href{http://arxiv.org/abs/1703.04584}{{\ttfamily arXiv:1703.04584 [hep-th]}}.

\bibitem{Anselmi:2017lia}
D.~Anselmi and M.~Piva, ``{Perturbative unitarity of Lee-Wick quantum field theory},'' \href{http://dx.doi.org/10.1103/PhysRevD.96.045009}{{\em Phys. Rev.} {\bfseries D96} no.~4, (2017) 045009},
\href{http://arxiv.org/abs/1703.05563}{{\ttfamily arXiv:1703.05563 [hep-th]}}.

\bibitem{Anselmi:2018kgz}
D.~Anselmi, ``{Fakeons And Lee-Wick Models},'' \href{http://dx.doi.org/10.1007/JHEP02(2018)141}{{\em JHEP} {\bfseries 02} (2018) 141},
\href{http://arxiv.org/abs/1801.00915}{{\ttfamily arXiv:1801.00915 [hep-th]}}.

\bibitem{Donoghue:2018lmc}
J.~F. Donoghue and G.~Menezes, ``{Massive poles in Lee-Wick quantum field theory},'' \href{http://dx.doi.org/10.1103/PhysRevD.99.065017}{{\em Phys. Rev. D} {\bfseries 99} no.~6, (2019) 065017}, \href{http://arxiv.org/abs/1812.03603}{{\ttfamily arXiv:1812.03603 [hep-th]}}.

\bibitem{Stelle:1976gc}
K.~S. Stelle, ``{Renormalization of Higher Derivative Quantum Gravity},'' \href{http://dx.doi.org/10.1103/PhysRevD.16.953}{{\em Phys. Rev. D} {\bfseries 16} (1977) 953--969}.

\bibitem{Tomboulis:1977jk}
E.~Tomboulis, ``{1/N Expansion and Renormalization in Quantum Gravity},'' \href{http://dx.doi.org/10.1016/0370-2693(77)90678-5}{{\em Phys. Lett. B} {\bfseries 70} (1977) 361--364}.

\bibitem{Julve:1978xn}
J.~Julve and M.~Tonin, ``{Quantum Gravity with Higher Derivative Terms},'' \href{http://dx.doi.org/10.1007/BF02748637}{{\em Nuovo Cim. B} {\bfseries 46} (1978) 137--152}.

\bibitem{Fradkin:1981iu}
E.~S. Fradkin and A.~A. Tseytlin, ``{Renormalizable asymptotically free quantum theory of gravity},'' \href{http://dx.doi.org/10.1016/0550-3213(82)90444-8}{{\em Nucl. Phys. B} {\bfseries 201} (1982) 469--491}.

\bibitem{Avramidi:1985ki}
I.~G. Avramidi and A.~O. Barvinsky, ``{ASYMPTOTIC FREEDOM IN HIGHER DERIVATIVE QUANTUM GRAVITY},'' \href{http://dx.doi.org/10.1016/0370-2693(85)90248-5}{{\em Phys. Lett. B} {\bfseries 159} (1985) 269--274}.

\bibitem{Antoniadis:1986tu}
I.~Antoniadis and E.~T. Tomboulis, ``{Gauge Invariance and Unitarity in Higher Derivative Quantum Gravity},'' \href{http://dx.doi.org/10.1103/PhysRevD.33.2756}{{\em Phys. Rev. D} {\bfseries 33} (1986) 2756}.

\bibitem{Johnston:1987ue}
D.~A. Johnston, ``{Sedentary Ghost Poles in Higher Derivative Gravity},'' \href{http://dx.doi.org/10.1016/0550-3213(88)90555-X}{{\em Nucl. Phys. B} {\bfseries 297} (1988) 721--732}.

\bibitem{Salvio:2014soa}
A.~Salvio and A.~Strumia, ``{Agravity},'' \href{http://dx.doi.org/10.1007/JHEP06(2014)080}{{\em JHEP} {\bfseries 06} (2014) 080}, \href{http://arxiv.org/abs/1403.4226}{{\ttfamily arXiv:1403.4226 [hep-ph]}}.

\bibitem{Salvio:2018crh}
A.~Salvio, ``{Quadratic Gravity},'' \href{http://dx.doi.org/10.3389/fphy.2018.00077}{{\em Front. in Phys.} {\bfseries 6} (2018) 77}, \href{http://arxiv.org/abs/1804.09944}{{\ttfamily arXiv:1804.09944 [hep-th]}}.

\bibitem{Donoghue:2018izj}
J.~F. Donoghue and G.~Menezes, ``{Gauge Assisted Quadratic Gravity: A Framework for UV Complete Quantum Gravity},'' \href{http://dx.doi.org/10.1103/PhysRevD.97.126005}{{\em Phys. Rev. D} {\bfseries 97} no.~12, (2018) 126005}, \href{http://arxiv.org/abs/1804.04980}{{\ttfamily arXiv:1804.04980 [hep-th]}}.

\bibitem{Donoghue:2019ecz}
J.~F. Donoghue and G.~Menezes, ``{Arrow of Causality and Quantum Gravity},'' \href{http://dx.doi.org/10.1103/PhysRevLett.123.171601}{{\em Phys. Rev. Lett.} {\bfseries 123} no.~17, (2019) 171601}, \href{http://arxiv.org/abs/1908.04170}{{\ttfamily arXiv:1908.04170 [hep-th]}}.

\bibitem{Donoghue:2021meq}
J.~F. Donoghue and G.~Menezes, ``{Causality and gravity},'' \href{http://dx.doi.org/10.1007/JHEP11(2021)010}{{\em JHEP} {\bfseries 11} (2021) 010}, \href{http://arxiv.org/abs/2106.05912}{{\ttfamily arXiv:2106.05912 [hep-th]}}.

\bibitem{Anselmi:2017ygm}
D.~Anselmi, ``{On the quantum field theory of the gravitational interactions},'' \href{http://dx.doi.org/10.1007/JHEP06(2017)086}{{\em JHEP} {\bfseries 06} (2017) 086},
\href{http://arxiv.org/abs/1704.07728}{{\ttfamily arXiv:1704.07728 [hep-th]}}.

\bibitem{Anselmi:2018ibi}
D.~Anselmi and M.~Piva, ``{The Ultraviolet Behavior of Quantum Gravity},'' \href{http://dx.doi.org/10.1007/JHEP05(2018)027}{{\em JHEP} {\bfseries 05} (2018) 027},
\href{http://arxiv.org/abs/1803.07777}{{\ttfamily arXiv:1803.07777 [hep-th]}}.

\bibitem{Anselmi:2018tmf}
D.~Anselmi and M.~Piva, ``{Quantum Gravity, Fakeons And Microcausality},'' \href{http://dx.doi.org/10.1007/JHEP11(2018)021}{{\em JHEP} {\bfseries 11} (2018) 021},
\href{http://arxiv.org/abs/1806.03605}{{\ttfamily arXiv:1806.03605 [hep-th]}}.

\bibitem{Anselmi:2018bra}
D.~Anselmi, ``{Fakeons, Microcausality And The Classical Limit Of Quantum Gravity},'' \href{http://dx.doi.org/10.1088/1361-6382/ab04c8}{{\em Class. Quant. Grav.} {\bfseries 36} (2019) 065010},
\href{http://arxiv.org/abs/1809.05037}{{\ttfamily arXiv:1809.05037 [hep-th]}}.

\bibitem{Donoghue:2021cza}
J.~F. Donoghue and G.~Menezes, ``{On quadratic gravity},'' \href{http://dx.doi.org/10.1393/ncc/i2022-22026-7}{{\em Nuovo Cim. C} {\bfseries 45} no.~2, (2022) 26}, \href{http://arxiv.org/abs/2112.01974}{{\ttfamily arXiv:2112.01974 [hep-th]}}.

\bibitem{Holdom:2021hlo}
B.~Holdom, ``{Ultra-Planckian scattering from a QFT for gravity},'' \href{http://dx.doi.org/10.1103/PhysRevD.105.046008}{{\em Phys. Rev. D} {\bfseries 105} no.~4, (2022) 046008}, \href{http://arxiv.org/abs/2107.01727}{{\ttfamily arXiv:2107.01727 [hep-th]}}.

\bibitem{Holdom:2021oii}
B.~Holdom, ``{Photon-photon scattering from a UV-complete gravity QFT},'' \href{http://dx.doi.org/10.1007/JHEP04(2022)133}{{\em JHEP} {\bfseries 04} (2022) 133}, \href{http://arxiv.org/abs/2110.02246}{{\ttfamily arXiv:2110.02246 [hep-ph]}}.

\bibitem{Piva:2023bcf}
M.~Piva, ``{Higher-derivative quantum gravity with purely virtual particles: renormalizability and unitarity},'' \href{http://dx.doi.org/10.1140/epjp/s13360-023-04486-0}{{\em Eur. Phys. J. Plus} {\bfseries 138} no.~10, (2023) 876}, \href{http://arxiv.org/abs/2305.12549}{{\ttfamily arXiv:2305.12549 [hep-th]}}.

\bibitem{Buoninfante:2023ryt}
L.~Buoninfante, ``{Massless and partially massless limits in Quadratic Gravity},'' \href{http://dx.doi.org/10.1007/JHEP12(2023)111}{{\em JHEP} {\bfseries 12} (2023) 111}, \href{http://arxiv.org/abs/2308.11324}{{\ttfamily arXiv:2308.11324 [hep-th]}}.

\bibitem{Buccio:2024hys}
D.~Buccio, J.~F. Donoghue, G.~Menezes, and R.~Percacci, ``{Physical Running of Couplings in Quadratic Gravity},'' \href{http://dx.doi.org/10.1103/PhysRevLett.133.021604}{{\em Phys. Rev. Lett.} {\bfseries 133} no.~2, (2024) 021604}, \href{http://arxiv.org/abs/2403.02397}{{\ttfamily arXiv:2403.02397 [hep-th]}}.

\bibitem{Ostrogradsky:1850fid}
M.~Ostrogradsky, ``{Memoires sur les equations differentielles, relatives au probleme des isoperimetres},''
{\em Mem. Acad. St. Petersbourg} {\bfseries 6} no.~4, (1850) 385--517.

\bibitem{Woodard:2015zca}
R.~P. Woodard, ``{Ostrogradsky's theorem on Hamiltonian instability},'' \href{http://dx.doi.org/10.4249/scholarpedia.32243}{{\em Scholarpedia} {\bfseries 10} no.~8, (2015) 32243},
\href{http://arxiv.org/abs/1506.02210}{{\ttfamily arXiv:1506.02210 [hep-th]}}.

\bibitem{Buoninfante:2022ykf}
L.~Buoninfante and K.~S. Kumar, ``{Quantum gravity, higher derivatives and nonlocality},'' \href{http://dx.doi.org/10.1393/ncc/i2022-22025-8}{{\em Nuovo Cim. C} {\bfseries 45} no.~2, (2022) 25}.

\bibitem{Kuntz:2024rzu}
J.~Kuntz, ``{Unitarity through PT symmetry in Quantum Quadratic Gravity},'' \href{http://arxiv.org/abs/2410.08278}{{\ttfamily arXiv:2410.08278 [hep-th]}}.

\bibitem{Deffayet:2021nnt}
C.~Deffayet, S.~Mukohyama, and A.~Vikman, ``{Ghosts without Runaway Instabilities},'' \href{http://dx.doi.org/10.1103/PhysRevLett.128.041301}{{\em Phys. Rev. Lett.} {\bfseries 128} no.~4, (2022) 041301}, \href{http://arxiv.org/abs/2108.06294}{{\ttfamily arXiv:2108.06294 [gr-qc]}}.

\bibitem{Deffayet:2023wdg}
C.~Deffayet, A.~Held, S.~Mukohyama, and A.~Vikman, ``{Global and local stability for ghosts coupled to positive energy degrees of freedom},'' \href{http://dx.doi.org/10.1088/1475-7516/2023/11/031}{{\em JCAP} {\bfseries 11} (2023) 031}, \href{http://arxiv.org/abs/2305.09631}{{\ttfamily arXiv:2305.09631 [gr-qc]}}.

\bibitem{ErrastiDiez:2024hfq}
V.~Errasti~D\'\i{}ez, J.~Gaset~Rif\`a, and G.~Staudt, ``{Foundations of ghost stability},'' \href{http://arxiv.org/abs/2408.16832}{{\ttfamily arXiv:2408.16832 [hep-th]}}.

\bibitem{Landshoff:1963nzy}
P.~V. Landshoff, ``{Poles and thresholds and unstable particles},'' \href{http://dx.doi.org/10.1007/bf02806056}{{\em Nuovo Cim.} {\bfseries 28} no.~1, (1963) 123--131}.

\bibitem{Eden:1966dnq}
R.~J. Eden, P.~V. Landshoff, D.~I. Olive, and J.~C. Polkinghorne, {\em {The analytic S-matrix}}.
\newblock Cambridge Univ. Press, Cambridge, 1966.

\bibitem{Brown:1992db}
L.~S. Brown, {\em {Quantum field theory}}.
\newblock Cambridge University Press, 7, 1994.

\bibitem{Coleman:2018mew}
S.~Coleman, \href{http://dx.doi.org/10.1142/9371}{{\em {Lectures of Sidney Coleman on Quantum Field Theory}}}.
\newblock WSP, Hackensack, 12, 2018.

\bibitem{Veltman:1963th}
M.~J.~G. Veltman, ``{Unitarity and causality in a renormalizable field theory with unstable particles},'' \href{http://dx.doi.org/10.1016/S0031-8914(63)80277-3}{{\em Physica} {\bfseries 29} (1963) 186--207}.

\bibitem{Donoghue:2019fcb}
J.~F. Donoghue and G.~Menezes, ``{Unitarity, stability and loops of unstable ghosts},'' \href{http://dx.doi.org/10.1103/PhysRevD.100.105006}{{\em Phys. Rev. D} {\bfseries 100} no.~10, (2019) 105006}, \href{http://arxiv.org/abs/1908.02416}{{\ttfamily arXiv:1908.02416 [hep-th]}}.

\bibitem{Kubo:2023lpz}
J.~Kubo and T.~Kugo, ``{Unitarity violation in field theories of Lee\textendash{}Wick\textquoteright{}s complex ghost},'' \href{http://dx.doi.org/10.1093/ptep/ptad143}{{\em PTEP} {\bfseries 2023} no.~12, (2023) 123B02}, \href{http://arxiv.org/abs/2308.09006}{{\ttfamily arXiv:2308.09006 [hep-th]}}.

\bibitem{Kubo:2024ysu}
J.~Kubo and T.~Kugo, ``{Anti-Instability of Complex Ghost},'' \href{http://dx.doi.org/10.1093/ptep/ptae053}{{\em PTEP} {\bfseries 2024} no.~5, (2024) 053B01}, \href{http://arxiv.org/abs/2402.15956}{{\ttfamily arXiv:2402.15956 [hep-th]}}.

\bibitem{Anselmi:2021hab}
D.~Anselmi, ``{Diagrammar of physical and fake particles and spectral optical theorem},'' \href{http://dx.doi.org/10.1007/JHEP11(2021)030}{{\em JHEP} {\bfseries 11} (2021) 030}, \href{http://arxiv.org/abs/2109.06889}{{\ttfamily arXiv:2109.06889 [hep-th]}}.

\bibitem{Holdom:2024cfq}
B.~Holdom, ``{UV-complete 4-derivative scalar field theory},'' \href{http://dx.doi.org/10.1016/j.nuclphysb.2024.116472}{{\em Nucl. Phys. B} {\bfseries 1000} (2024) 116472}, \href{http://arxiv.org/abs/2402.09223}{{\ttfamily arXiv:2402.09223 [hep-th]}}.

\bibitem{tHooft:1973wag}
G.~'t~Hooft and M.~J.~G. Veltman, ``{DIAGRAMMAR},'' \href{http://dx.doi.org/10.1007/978-1-4684-2826-1_5}{{\em NATO Sci. Ser. B} {\bfseries 4} (1974) 177--322}.

\bibitem{Anselmi:2020lfx}
D.~Anselmi, ``{Dressed propagators, fakeon self-energy and peak uncertainty},'' \href{http://dx.doi.org/10.1007/JHEP06(2022)058}{{\em JHEP} {\bfseries 22} (2020) 058}, \href{http://arxiv.org/abs/2201.00832}{{\ttfamily arXiv:2201.00832 [hep-ph]}}.

\bibitem{Holdom:2023usn}
B.~Holdom, ``{Running couplings and unitarity in a 4-derivative scalar field theory},'' \href{http://dx.doi.org/10.1016/j.physletb.2023.138023}{{\em Phys. Lett. B} {\bfseries 843} (2023) 138023}, \href{http://arxiv.org/abs/2303.06723}{{\ttfamily arXiv:2303.06723 [hep-th]}}.

\bibitem{Bender:2007wu}
C.~M. Bender and P.~D. Mannheim, ``{No-ghost theorem for the fourth-order derivative Pais-Uhlenbeck oscillator model},'' \href{http://dx.doi.org/10.1103/PhysRevLett.100.110402}{{\em Phys. Rev. Lett.} {\bfseries 100} (2008) 110402}, \href{http://arxiv.org/abs/0706.0207}{{\ttfamily arXiv:0706.0207 [hep-th]}}.

\bibitem{Bender:2008gh}
C.~M. Bender and P.~D. Mannheim, ``{Exactly solvable PT-symmetric Hamiltonian having no Hermitian counterpart},'' \href{http://dx.doi.org/10.1103/PhysRevD.78.025022}{{\em Phys. Rev. D} {\bfseries 78} (2008) 025022}, \href{http://arxiv.org/abs/0804.4190}{{\ttfamily arXiv:0804.4190 [hep-th]}}.

\bibitem{Mannheim:2009zj}
P.~D. Mannheim, ``{PT symmetry as a necessary and sufficient condition for unitary time evolution},'' \href{http://dx.doi.org/10.1098/rsta.2012.0060}{{\em Phil. Trans. Roy. Soc. Lond. A} {\bfseries 371} (2013) 20120060}, \href{http://arxiv.org/abs/0912.2635}{{\ttfamily arXiv:0912.2635 [hep-th]}}.

\bibitem{Salvio:2015gsi}
A.~Salvio and A.~Strumia, ``{Quantum mechanics of 4-derivative theories},'' \href{http://dx.doi.org/10.1140/epjc/s10052-016-4079-8}{{\em Eur. Phys. J. C} {\bfseries 76} no.~4, (2016) 227}, \href{http://arxiv.org/abs/1512.01237}{{\ttfamily arXiv:1512.01237 [hep-th]}}.

\bibitem{Strumia:2017dvt}
A.~Strumia, ``{Interpretation of quantum mechanics with indefinite norm},'' \href{http://dx.doi.org/10.3390/physics1010003}{{\em MDPI Physics} {\bfseries 1} no.~1, (2019) 17--32}, \href{http://arxiv.org/abs/1709.04925}{{\ttfamily arXiv:1709.04925 [quant-ph]}}.

\bibitem{Holdom:2024onr}
B.~Holdom, ``{Making sense of ghosts},'' \href{http://dx.doi.org/10.1016/j.nuclphysb.2024.116696}{{\em Nucl. Phys. B} {\bfseries 1008} (2024) 116696}, \href{http://arxiv.org/abs/2408.04089}{{\ttfamily arXiv:2408.04089 [hep-th]}}.

\bibitem{Asorey:2024mkb}
M.~Asorey, G.~Krein, and I.~L. Shapiro, ``{Bound states of massive complex ghosts in superrenormalizable quantum gravity theories},'' \href{http://arxiv.org/abs/2408.16514}{{\ttfamily arXiv:2408.16514 [gr-qc]}}.

\bibitem{Shapiro:2015uxa}
I.~L. Shapiro, ``{Counting ghosts in the \textquotedblleft{}ghost-free\textquotedblright{} non-local gravity},'' \href{http://dx.doi.org/10.1016/j.physletb.2015.03.037}{{\em Phys. Lett. B} {\bfseries 744} (2015) 67--73}, \href{http://arxiv.org/abs/1502.00106}{{\ttfamily arXiv:1502.00106 [hep-th]}}.

\bibitem{Buoninfante:2018mre}
L.~Buoninfante, G.~Lambiase, and A.~Mazumdar, ``{Ghost-free infinite derivative quantum field theory},'' \href{http://dx.doi.org/10.1016/j.nuclphysb.2019.114646}{{\em Nucl. Phys. B} {\bfseries 944} (2019) 114646}, \href{http://arxiv.org/abs/1805.03559}{{\ttfamily arXiv:1805.03559 [hep-th]}}.

\bibitem{Briscese:2024tvc}
F.~Briscese, G.~Calcagni, L.~Modesto, and G.~Nardelli, ``{Form factors, spectral and K\"all\'en-Lehmann representation in nonlocal quantum gravity},'' \href{http://dx.doi.org/10.1007/JHEP08(2024)204}{{\em JHEP} {\bfseries 08} (2024) 204}, \href{http://arxiv.org/abs/2405.14056}{{\ttfamily arXiv:2405.14056 [hep-th]}}.

\bibitem{Platania:2020knd}
A.~Platania and C.~Wetterich, ``{Non-perturbative unitarity and fictitious ghosts in quantum gravity},'' \href{http://dx.doi.org/10.1016/j.physletb.2020.135911}{{\em Phys. Lett. B} {\bfseries 811} (2020) 135911}, \href{http://arxiv.org/abs/2009.06637}{{\ttfamily arXiv:2009.06637 [hep-th]}}.

\bibitem{Platania:2022gtt}
A.~Platania, ``{Causality, unitarity and stability in quantum gravity: a non-perturbative perspective},'' \href{http://dx.doi.org/10.1007/JHEP09(2022)167}{{\em JHEP} {\bfseries 09} (2022) 167}, \href{http://arxiv.org/abs/2206.04072}{{\ttfamily arXiv:2206.04072 [hep-th]}}.

\bibitem{Anselmi:2022toe}
D.~Anselmi, ``{Purely virtual particles versus Lee-Wick ghosts: Physical Pauli-Villars fields, finite QED, and quantum gravity},'' \href{http://dx.doi.org/10.1103/PhysRevD.105.125017}{{\em Phys. Rev. D} {\bfseries 105} no.~12, (2022) 125017}, \href{http://arxiv.org/abs/2202.10483}{{\ttfamily arXiv:2202.10483 [hep-th]}}.

\bibitem{Anselmi:2023wjx}
D.~Anselmi, ``{Propagators and widths of physical and purely virtual particles in a finite interval of time},'' \href{http://dx.doi.org/10.1007/JHEP07(2023)099}{{\em JHEP} {\bfseries 07} (2023) 099}, \href{http://arxiv.org/abs/2304.07643}{{\ttfamily arXiv:2304.07643 [hep-ph]}}.

\bibitem{Takahashi:1996zn}
Y.~Takahashi and H.~Umezawa, ``{Thermo field dynamics},'' \href{http://dx.doi.org/10.1142/S0217979296000817}{{\em Int. J. Mod. Phys. B} {\bfseries 10} (1996) 1755--1805}.

\bibitem{DeFilippo:1977bk}
S.~De~Filippo and G.~Vitiello, ``{Vacuum Structure for Unstable Particles},'' \href{http://dx.doi.org/10.1007/BF02746504}{{\em Lett. Nuovo Cim.} {\bfseries 19} (1977) 92}.

\bibitem{Rabuffo:1977va}
I.~Rabuffo and G.~Vitiello, ``{Vacuum Structure for Indefinite Metric Quantum Field Theory},'' \href{http://dx.doi.org/10.1007/BF02812981}{{\em Nuovo Cim. A} {\bfseries 44} (1978) 401}.

\end{thebibliography}\endgroup


\end{document}